# Graphite and Hexagonal Boron-Nitride Possess the Same Interlayer Distance. Why?

Oded Hod

School of Chemistry, The Sackler Faculty of Exact Sciences, Tel Aviv University, Tel Aviv 69978, Israel

Graphite and hexagonal boron nitride ($h$-BN) are two prominent members of the family of layered materials possessing a hexagonal lattice structure. While graphite has non-polar homo-nuclear C-C intra-layer bonds, $h$-BN presents highly polar B-N bonds resulting in different optimal stacking modes of the two materials in bulk form. Furthermore, the static polarizabilities of the constituent atoms considerably differ from each other suggesting large differences in the dispersive component of the interlayer bonding. Despite these major differences both materials present practically identical interlayer distances. To understand this finding, a comparative study of the nature of the interlayer bonding in both materials is presented. A full lattice sum of the interactions between the partially charged atomic centers in $h$-BN results in vanishingly small monopolar electrostatic contributions to the interlayer binding energy. Higher order electrostatic multipoles, exchange, and short-range correlation contributions are found to be very similar in both materials and to almost completely cancel out by the Pauli repulsions at physically relevant interlayer distances resulting in a marginal effective contribution to the interlayer binding. Further analysis of the dispersive energy term reveals that despite the large differences in the individual atomic polarizabilities the hetero-atomic B-N $C_6$ coefficient is very similar to the homo-atomic C-C coefficient in the hexagonal bulk form resulting in very similar dispersive contribution to the interlayer binding. The overall binding energy curves of both materials are thus very similar predicting practically the same interlayer distance and very similar binding energies. The conclusions drawn here regarding the role of monopolar electrostatic interactions for the interlayer binding of $h$-BN are of general nature and are expected to hold true for many other polar layered systems.

Layered materials are playing a central role in a variety of key scientific fields including nano-scale materials science, condensed matter physics, molecular electronics and spintronics, tribology, and chemistry. While the intra-layer interactions are often well characterized and dominated by covalent bonding, the inter-layer interactions are determined by a delicate balance between dispersion forces, electrostatic interactions and Pauli repulsions. Understanding the relative contribution of each of these interactions to the inter-layer binding is therefore essential for the characterization of their mechanical, electronic, and electromechanical properties and for the design of new materials with desired functionality.[1-7]

In recent years, the most prominent member of the family of layered materials has been graphene[9-12] which serves as a building block for few layered graphene and graphite as well as for single- and multi-walled carbon nanotubes.[13] Here, each layer is an atomically thin hexagonal sheet of $SP^2$ bonded carbon atoms, where the unpaired $p_z$ electrons on each atomic site join to form a collective π system turning the material into a semi-metal. The non-polar nature of the homo-nuclear carbon-carbon bonds results in zero formal charges on each atomic center thus excluding monopolar electrostatic contributions to the interlayer binding. This leaves higher electrostatic multipoles, dispersion interactions, and Pauli repulsion as the dominant factors governing the stacking and registry of the layered structure.[14-26] Here, the complex interplay between these factors dictates the equilibrium interlayer distance[16] and the optimal AB staking mode (see Fig. 1) where consecutive layers are shifted with respect to each other such that half of the carbon atoms of one layer reside above the hexagon centers of the adjacent layers.[6-7]

The inorganic analog of graphene, sometimes referred to as "white graphene",[27-29] is hexagonal boron nitride.[30-40] Structurally, a single layer of *h*-BN is very similar to a graphene sheet having a hexagonal backbone where each couple of bonded carbon atoms is replaced by a boron-nitride pair. Furthermore, the two materials are isoelectronic. Nevertheless, due to the electronegativity differences between the boron and the nitrogen atoms the π electrons tend to localize around the nitrogen atomic centers[41-44] thus forming an insulating material. Furthermore, the polarity of the B-N bond results in formal charges around the atomic centers thus allowing for monopolar inter-layer electrostatic interactions to join higher electrostatic multipoles, dispersion interactions, and Pauli repulsion in dictating the nature of the interlayer binding. This, in turn, stabilizes the AA' stacking mode (see Fig. 1) where a boron atom bearing a partial positive charge in one layer resides on top of the oppositely charged nitrogen atoms on the adjacent layers.

Based on the above considerations, one may generally deduce that electrostatic interactions between partially charged atomic centers may play a crucial role in the interlayer binding of polar layered materials.[41] Specifically, the electrostatic attractions between the oppositely charged atomic centers in adjacent *h*-BN layers are expected to result in a considerably shorter interlayer distance than that measured in graphite. Nevertheless, the interlayer distances in graphite (3.33-3.35 Å)[45-47] and in *h*-BN (3.30-3.33 Å)[48-53] are essentially the same suggesting that monopolar electrostatic interactions, which exist in *h*-BN and are absent in graphite, have little effect on the interlayer binding. This is consistent with a recent study showing that van der Waals (vdW) forces, rather than electrostatic interactions, are responsible for anchoring the *h*-BN layers at the appropriate interlayer distance.[7] Further support for this argument is found when comparing the optimal AA' stacking mode with the $AB_1$ stacking mode where the partially positively charged boron atomic sites are eclipsed and the nitrogen atoms reside atop hexagon centers in adjacent layers (see Fig. 1). From a naive electrostatic viewpoint one would expect the AA' mode, where opposite charges reside atop of each other, to be considerably lower in energy than the $AB_1$ mode whereas according to advanced *ab-initio* calculations the later is found to be only 0.875-2.0 meV/atom higher in energy.[7,29,54] Furthermore, when comparing to the $AB_2$ stacking mode (see Fig. 1), which in terms of monopolar contributions is electrostatically equivalent to the $AB_1$ mode, its total energy is higher by as much as 6.5-12.0 meV/atom than both the AA' and the $AB_1$ modes.[7,54-55] This may be related to enhanced Pauli repulsions between the more delocalized overlapping electron clouds of the nitrogen atoms.[4,29,54,56-59]

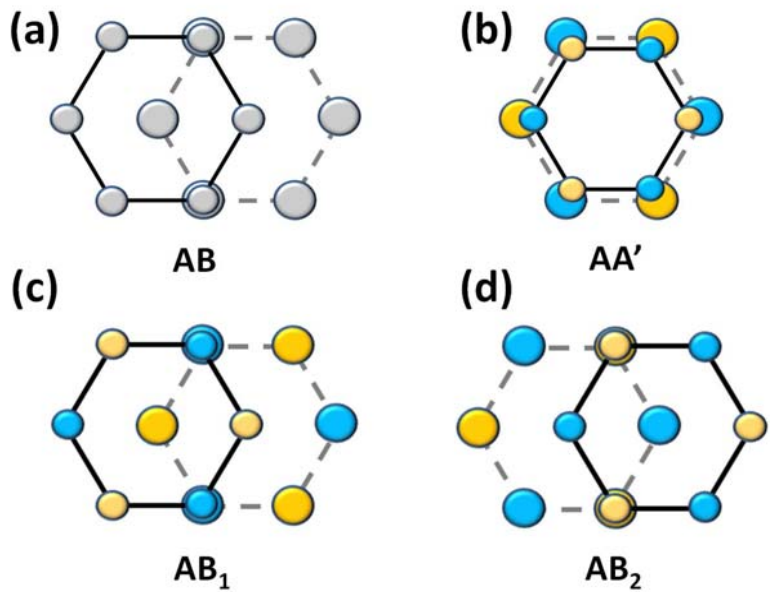

Figure 1: High symmetry stacking modes in hexagonal lattices. (a) The optimal AB stacking of graphite has a carbon atom in a given layer residing atop of the center of a hexagon in the adjacent layers. (b) The optimal AA' stacking mode of *h*-BN has a partially negatively charged nitrogen atom in one layer residing on-top of a partially positively charged boron atom in the adjacent layers. This configuration minimizes the electrostatic energy. (c) The meta-stable $AB_1$ stacking mode of *h*-BN has eclipsed boron atom positions whereas the nitrogen atoms appear on-top of hexagon centers of adjacent layers. (d) The non-stable $AB_2$ stacking mode of *h*-BN has eclipsed nitrogen atom positions whereas the boron atoms appear on-top of hexagon centers of adjacent layers. Lower(upper) layer hexagons are indicated by larger(smaller) circles representing the atoms and dashed(full) lines representing $SP^2$ covalent bonds. Blue(orange) circles represent boron(nitrogen) atoms. Electrostatically, the $AB_1$ and $AB_2$ stacking modes should be energetically equivalent, however due to the vanishing monopolar electrostatic interactions between adjacent layers and enhanced Pauli repulsions between eclipsed nitrogen centers the $AB_1$ configuration is close in total energy to the AA' stacking mode whereas the $AB_2$ is a non-stable high energy mode.

To add to the puzzle, even if one accepts that monopolar electrostatic interactions do not contribute to the binding in the polar *h*-BN system, the differences in spatial distribution of the charge densities in both systems would suggest that the contribution of higher electrostatic multipoles would be different in both materials. Furthermore, the large differences in the values of the static polarizabilities of the boron, carbon, and nitrogen atoms suggest that the dispersive contribution to the binding would behave differently in both materials.

Several questions thus arise: Why is the effect of monopolar electrostatic interactions on the interlayer binding of *h*-BN negligible? Why is the effect of higher electrostatic multipoles in both graphene and *h*-BN similar? Why is the dispersive attraction similar in both materials? And more generally, is the fact that the interlayer distances of graphite and *h*-BN are so similar a mere coincidence or an effect of a more generic nature?

To answer these questions we start by considering the marginal effect of monopolar electrostatic contributions on the binding energy of *h*-BN. Here, the answer lies in the long-range nature of the Coulomb interactions. Our intuition for enhanced electrostatic binding in *h*-BN stems from the attraction of oppositely charged boron and nitrogen atoms residing opposite to each other on adjacent layers at the optimal AA' stacking mode. Nevertheless, the interlayer Coulomb interactions between laterally shifted atomic sites are non-negligible and must be appropriately taken into account. Specifically, as the lateral distance $r$ from a test charge is increased, the Coulomb interaction decays as $\alpha/r$ where $\alpha$ is the effective partial charge on each atomic site. However, the number of atomic sites interacting with the test charge at the given lateral distance $r$ is approximately proportional to the circumference of a ring of radius $r$ and thus increases linearly with the distance. Thus, as previously discussed,[4-5,60-70] in order to map the monopolar electrostatic potential above an infinite *h*-BN layer, it is necessary to perform a full lattice sum over all partially charged lattice sites within the sheet. This sum is given by the following general expression:

$$\varphi(\vec{r}) = \sum_{i=1}^{d} \sum_{n=-\infty}^{\infty} \sum_{m=-\infty}^{\infty} \frac{q_i}{\left[(x - x_i - nT_1^x - mT_2^x)^2 + (y - y_i - nT_1^y - mT_2^y)^2 + (z - z_i)^2\right]^{1/2}} \quad (1)$$

where, $\varphi(\vec{r})$ is the electrostatic potential, in atomic units, at point $\vec{r} = (x, y, z)$ in space due to $q_{i=1,\ldots,d}$ charges located at points $\vec{r}_i = (x_i, y_i, z_i)$ within the two-dimensional periodic unit cell with lattice vectors $\vec{T}_1 = (T_1^x, T_1^y)$ and $\vec{T}_2 = (T_2^x, T_2^y)$. We note that $\varphi(\vec{r})$ diverges when measured at the lattice sites. For simplicity we choose a rectangular unit cell (see right panel of Fig. 2) with lattice vectors $\vec{T}_1 = (\sqrt{3}, 0)a$ and $\vec{T}_2 = (0,3)a$, $a$=1.446 Å being the B-N bond length, atomic positions $\vec{r}_1 = (0,0,0)$, $\vec{r}_2 = (0,1,0)a$, $\vec{r}_3 = \frac{1}{2}(\sqrt{3}, 3, 0)a$, $\vec{r}_4 = \frac{1}{2}(\sqrt{3}, 5, 0)a$, and charges $q_1 = -q_2 = q_3 = -q_4 = \delta$ where $\delta = -0.5$.[7,60] Unfortunately, for charge neutral unit-cells ($\sum_{i=1}^{d} q_i$), the sum appearing in Eq. (1) is conditionally convergent and therefore challenging to evaluate using direct summation. An elegant way to circumvent this problem was proposed by Ewald where the conditionally convergent lattice sum is converted into two absolutely converging sums one in real space and the other in reciprocal space.[71-72] Using this technique (see supporting material for a detailed derivation) one is able to efficiently calculate the electrostatic potential at any point above the two dimensional *h*-BN lattice.

To study the electrostatic contribution at the optimal AA' stacking mode the electrostatic potential above a nitrogen atomic site is plotted as a function of distance from the *h*-BN layer. In the left panel of Fig. 2 the full lattice-sum results are compared to the electrostatic potential produced by an isolated partially charged nitrogen atomic center. Clearly, the collective electrostatic potential decays exponentially (see supporting material) and much faster than $-\delta/r$ becoming extremely small at the equilibrium interlayer distance in agreement with similar results obtained by Green *et al.*[5] At shorter distances Pauli repulsions become dominant and prevent the layers from approaching each other thus rendering the region, where monopolar electrostatic interactions contributions become substantial for binding, physically irrelevant. At the optimal AA' stacking mode with interlayer distance set to 3.33 Å the calculated monopolar electrostatic potential energy is $8.4 \cdot 10^{-4}$ eV/atom which is merely a negligible fraction of the total bilayer binding energy in the presence of vdW interactions calculated to be 26.0-38.1 meV/atom.[7,24] A full map of the monopolar electrostatic potential 3.33 Å above the *h*-BN surface is presented in the right panel of Fig. 2. Due to symmetry considerations the potential vanishes identically above the centers of the hexagons and above the centers of the B-N bonds regardless of the distance from the surface. At other positions along the surface the potential is non-zero and preserves the hexagonal lattice symmetry with values not exceeding $8.4 \cdot 10^{-4}$ eV/atom.

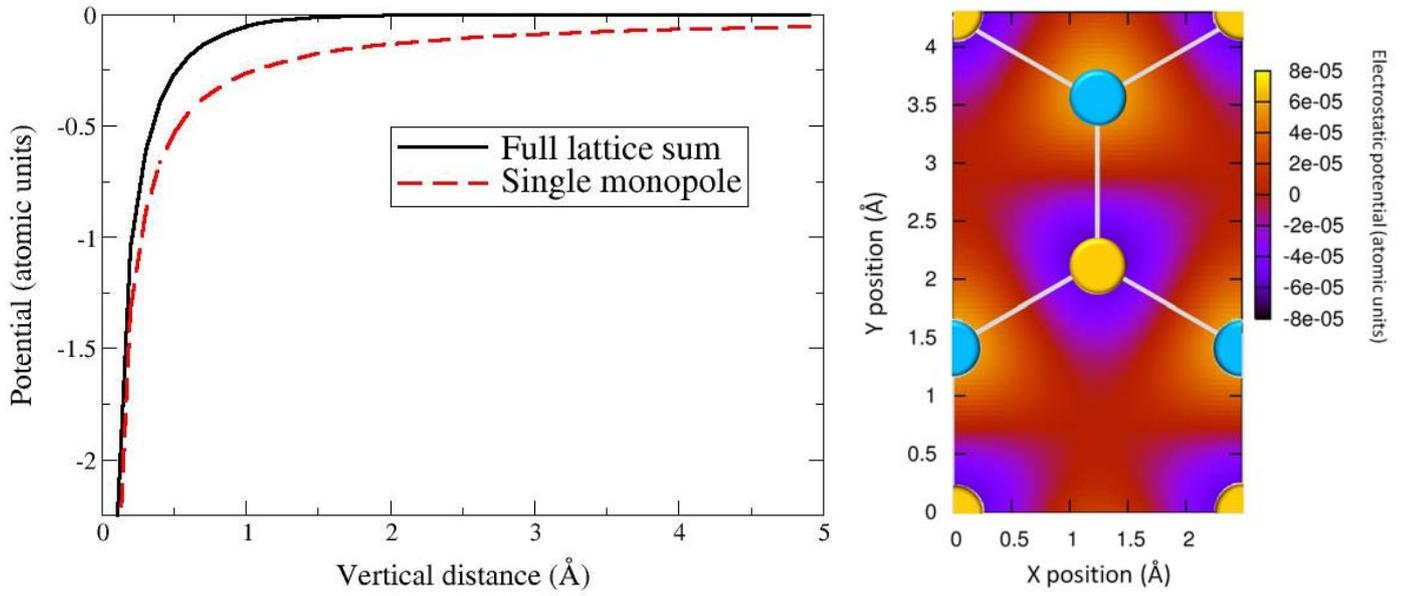

Figure 2: Electrostatic potential (atomic units) above an *h*-BN surface. Left panel: Electrostatic potential above a partially negatively charged (-0.5 $e^-$) nitrogen site as a function of vertical distance from the plane of the *h*-BN layer (Solid black curve) calculated using Eq. (1). For comparison purposes the potential above a corresponding partially charged isolated nitrogen atom is presented by the dashed red line. Right panel: Full electrostatic potential surface 3.33 Å above the *h*-BN layer calculated using Eq. (1). Boron(nitrogen) atomic positions are represented by blue (orange) circles. For the optimal AA' stacking mode at the equilibrium interlayer distance of 3.33 Å the lattice summed electrostatic potential becomes extremely small (left panel). Due to symmetry considerations the electrostatic potential above the center of the hexagon and above the center of a B-N bond vanishes identically (right panel).

Based on the above considerations it is now clear that due to the long-range nature of the Coulomb interactions the overall monopolar electrostatic attraction between partially charged atomic centers has only a marginal effect on the interlayer binding in *h*-BN. One may therefore conclude that the main electrostatic contribution to the interlayer binding in both graphene and *h*-BN comes from higher order multipoles.[4] It is therefore tempting to assume that these contributions would be very similar in both materials due to the almost identical intra-layer hexagonal lattice structure they possess. This, however, is not trivial as both the optimal stacking mode and the overall density profile in the two materials are quite different. Thus, in order to gain better understanding of the separate role of higher order electrostatic multipoles for the interlayer binding, density functional theory (DFT) based binding energy calculations for bilayer graphene and *h*-BN have been performed. As will be discussed below, these calculations also allow for probing the role of Pauli repulsions in preventing the layers from sticking together.[16] In order to avoid ambiguities in the definition of the different components of the total energy resulting from the lattice sums performed in periodic boundary conditions calculations a set of finite-sized bilayer clusters with hexagonal symmetry and increasing diameter has been chosen. For the *h*-BN system zigzag edged hydrogen terminated hexagonal clusters have been considered (see right panel of Fig. 3), test calculations with armchair *h*-BN clusters revealed similar results to those obtained with the zigzag clusters (see supporting material). In order to prevent the occurrence of edge states in the bilayer graphene system,[12,73-77] hydrogen terminated armchair graphene dimmers have been considered

(see left panel of Fig. 3). Each hexagonal cluster was cut out of the relevant pristine periodic layer with C-C and B-N bond lengths of 1.420 Å and 1.446 Å, respectively. The bare edges were hydrogen terminated with the benzene C-H, and borazine B-H and N-H bond lengths of 1.101 Å, 1.200 Å, and 1.020 Å, respectively. The individual flakes were then appropriately combined to form a finite sized AB stacked graphene dimmer and AA' stacked *h*-BN dimmer. No geometry optimization was performed. The cluster size was increased until edge effects on the calculated binding energies became marginal (see supporting material). All calculations were carried out using the Gaussian 09 suite of programs[78] with the double-$\zeta$ polarized 6-31G** Gaussian basis set[79] utilizing the counter-poise correction[80-81] to eliminate possible basis set super position errors. Tests for convergence of the results with respect to the basis sets were performed for the smaller flakes indicating convergence of the total binding energy down to ~1 meV/atom at physically relevant interlayer separations (see supporting material).

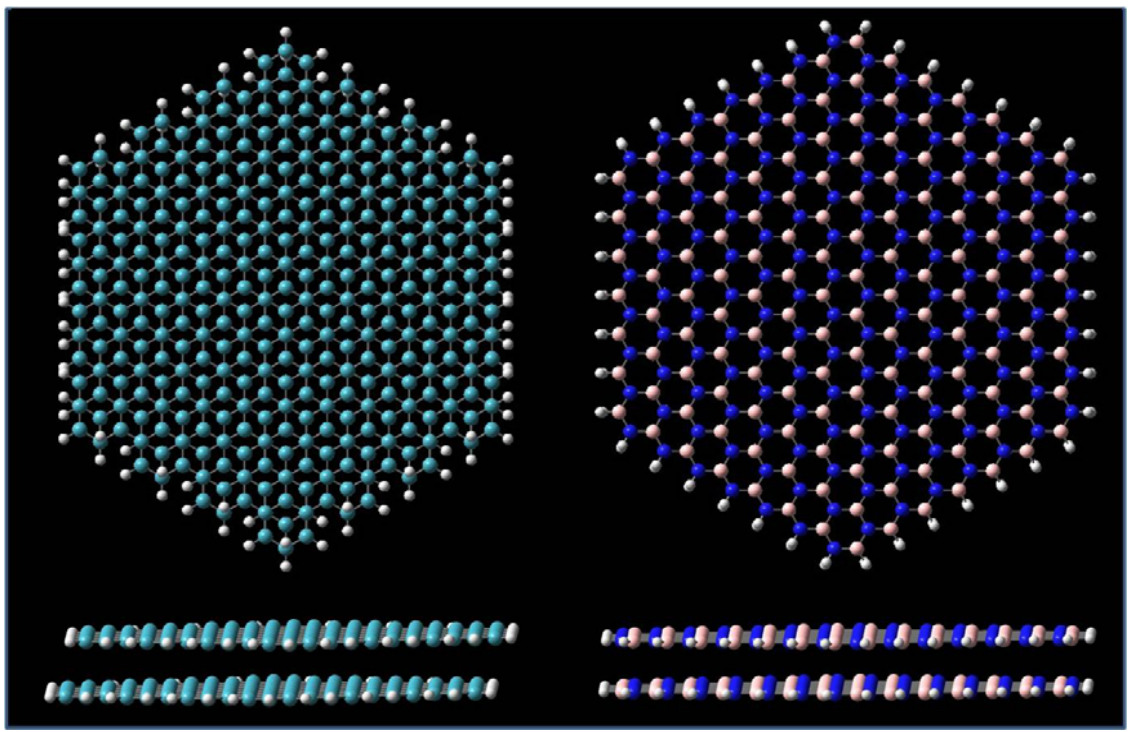

Figure 3: Top- (upper panels) and side- (lower panels) views of the largest bilayer-graphene armchair flakes (left) and bilayer *h*-BN zigzag flakes (right) used in the present study. The graphene system consists of a total of 528 atoms and the h-BN system has a total of 672 atoms. Cyan, blue, pink and grey spheres represent carbon, nitrogen, boron, and hydrogen atoms, respectively.

Fig. 4 presents the dependence of different components of the total energy on the interlayer distance in graphene and *h*-BN. Here, $E_{El}$ is the sum of classical electrostatic contributions (nuclear-nuclear repulsion, electron-nuclear attraction and the Hartree term), $E_{xc}$ is the sum of exchange and correlation DFT contributions, $E_k$ is the kinetic energy term, and $E_T$ – the total energy. Two exchange-correlation density functional approximations are considered:[82] the generalized gradient corrected PBE functional[83] representing semi-local functionals and the hybrid B3LYP functional[84] aimed at partly eliminating the self interaction error appearing in semi-local functionals and regaining some of the correct long-range exchange behavior which is relevant for the present study. Both functional approximations lack the proper treatment of long-range correlation effects responsible for dispersive vdW interactions and are therefore limited to the description of classical electrostatic, exchange, short-range (SR) correlation, and Pauli repulsions effects on the interlayer binding.

As can be seen, both functional approximations predict that higher-order electrostatic multipoles contributions (red squares) are much larger than the $h$-BN monopole energy (brown 'x' marks) at physically relevant interlayer distances of the two materials. Nevertheless, the combined electrostatic, exchange, and SR-correlation (green diamonds) contributions to the total binding energy at these distances are almost completely canceled out by the Pauli repulsions manifested in the kinetic energy term (blue triangles).[16] As a result, the total binding energy curves (black circles) calculated by both functionals which, as described above, lack the dispersive component, are completely non-bonding for graphene and very weakly bonding for $h$-BN. This is consistent with recently reported results for graphite[85] and molecular graphene derivatives adsorbed on graphene.[16]

The PBE calculations suggest that while the dependence of the exchange-SR-correlation contributions on the interlayer distance in both materials is very similar the electrostatic and kinetic energy terms of graphene and $h$-BN behave quite differently. This implies that the similarity of the total (vdW lacking) binding energy curves of the two materials results from a coincidental cancelation of the different terms. The B3LYP results reveal a different picture where the interlayer distance dependence of all energy components in both materials are very similar (with minor deviations between the kinetic energy terms) thus systematically producing very similar (vdW lacking) binding energy curves.

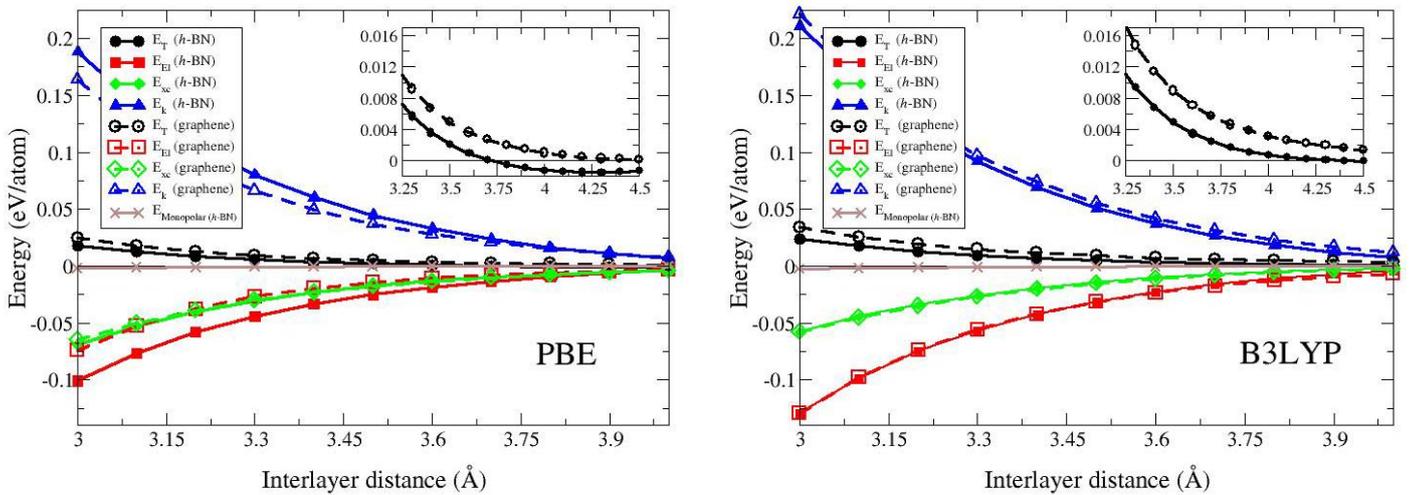

Figure 4: Dependence of the total (black circles), electrostatic (red squares), exchange-SR-correlation (green diamonds), and kinetic (blue triangles) energy components on the interlayer distance of bilayer $h$-BN (solid lines, full symbols) and bilayer graphene (dashed lines, open symbols) calculated using the PBE (left panel) and B3LYP (right panel) density functional approximations. Zero energy is defined as the value of the relevant component at infinite separation. Insets: zoom on the interlayer dependence of the calculated total energies. PBE results for bilayer graphene($h$-BN) were obtained using the 528(672) atom cluster presented in the left(right) panel of Fig. 3. B3LYP results for bilayer graphene and $h$-BN were obtained using 528 and 240 atom clusters, respectively (see supporting material).

The analysis presented above establishes the fact that electrostatic interactions between the partially charged atomic cores in *h*-BN, which are absent in graphene, have minor contribution to the interlayer binding due to the rapid decay of the potential into the vacuum above the layer. Furthermore, it shows that at physically relevant interlayer distances in graphene and *h*-BN the contributions of higher order electrostatic multipoles and exchange-SR-correlation contributions (which by themselves are quite significant) almost completely cancel out with the kinetic energy term manifesting the effect of Pauli repulsions. This suggests that vdW interactions are a crucial ingredient for anchoring the graphene and *h*-BN layers at their equilibrium interlayer distance.[7-8] Since the experimental interlayer distances in both systems are essentially the same one may deduce that the attractive vdW interactions in both systems are similar. As mentioned above, this conclusion is somewhat surprising in light of the different static polarizabilities presented by the carbon, boron, and nitrogen atoms.

In order to gain quantitative understanding regarding the role of vdW interactions for the interlayer binding in the two materials the $C_6/R^6$ leading dispersion term should be considered. To this end, the Tkatchenko-Scheffler vdW (TS-vdW) correction scheme to density functional theory may be used.[86-87] Here, the pair-wise bulk $C_6$ coefficients between atoms $A$ and $B$ are calculated using the following relation:[86]

$$C_{6,AB} = \frac{2 C_{6,A} C_{6,B}}{\frac{\alpha_B}{\alpha_A} C_{6,A} + \frac{\alpha_A}{\alpha_B} C_{6,B}} \quad (2)$$

where, $\alpha_i$ is the bulk static polarizability and $C_{6,i}$ is the homo-nuclear pair-wise bulk coefficient of atom $i = A, B$. The homo-nuclear bulk polarizabilities and coefficients can be obtained from the free atom values ($\alpha_i^0$ and $C_{6,i}^0$, respectively) via:

$$\alpha_i = \left(\frac{V_i^{eff}}{V_i^{free}}\right) \alpha_i^0 \quad ; \quad C_{6,i} = \left(\frac{V_i^{eff}}{V_i^{free}}\right)^2 C_{6,i}^0 \quad (3)$$

where, $V_i^{eff}$ is the effective volume of atom $i$ in the bulk system referenced to the free atom volume in vacuo, $V_i^{free}$. The relative effective volume, in turn, is estimated using the Hirshfeld partitioning scheme.[88]

The free-atom parameters may be obtained from the database presented by Chu and Dalgarno,[89] constructed using self-interaction corrected time dependent density functional theory calculations. Values for the relevant atoms are summarized in the table 1.

|  | C | B | N |
|---|---|---|---|
| $\alpha^0$ (a.u.)[89] | 12.0 | 21.0 | 7.4 |
| $C_6^0$ (a.u.)[89] | 46.6 | 99.5 | 24.2 |
| $V_i^{eff}/V_i^{free}$ [90] | 0.850 (graphite) | 0.811 (*h*-BN) | 0.879 (*h*-BN) |

Table 1: Values for the free atom dipole polarizabilities, $C_6$ coefficients, and relative effective Hirshfeld volumes of carbon, boron, and nitrogen atoms relevant for the present study.

Table 2 summarizes the numerical values for the pair-wise bulk (graphite and *h*-BN) $C_6$ coefficients obtained using Eqs. (2) and (3) with the parameters presented in table 1. At the optimal AA' stacking mode of *h*-BN and AB mode of graphite the most prominent $C_6$ contributions come from the eclipsed boron-nitrogen (in *h*-BN) and carbon-carbon (in graphite) atomic centers attraction on adjacent layers. As can be seen, despite the large differences between the C-C, B-B and N-N coefficients the C-C and B-N coefficient agree to within less than 2% indicating that indeed the vdW interactions in graphite and *h*-BN should be very similar in nature.

| B-B | N-N | C-C | B-N | C-B | C-N |
|---|---|---|---|---|---|
| 65.4 | 18.7 | 33.7 | 33.1 | 46.2 | 24.8 |

Table 2: Values (in Hartree*Bohr$^6$) for the pair-wise bulk $C_6$ coefficients obtained using Eqs. (2) and (3) and the parameters of table 1 for carbon, boron, and nitrogen atoms. Values relevant for the present study are shaded in gray.

To further investigate the vdW contribution beyond the eclipsed atoms interactions an analysis of the full vdW interaction scheme of the bilayer systems is presented in Fig. 5 where the *h*-BN bilayer is assumed to be at the AA' stacking mode and bilayer graphene at the AB mode. Different components of the vdW energy are considered separately. The term "mixed sub-lattice" interactions in *h*-BN refers to the vdW energy contribution of a single boron(nitrogen) atom in one *h*-BN layer with all nitrogen(boron) atoms in the other layer (marked as BN). In bilayer graphene this term refers to the interaction of a single carbon atom located at a given sub-lattice site of one graphene layer with all carbon atoms belonging to the other sub-lattice sites of the second graphene layer (marked as CC' or C'C). Respectively, the term "same sub-lattice" interactions in *h*-BN refers to the vdW contribution of a single boron(nitrogen) atom in one *h*-BN layer with all boron(nitrogen) atoms in the other layer (marked as BB or NN). In bilayer graphene this term refers to the interaction of a single carbon atom located in a given sub-lattice site of one graphene layer with all carbon atoms belonging to the same sub-lattice sites of the second graphene layer (marked as CC or C'C').

Panel (a) of Fig. 5 shows the vdW energy contribution of the mixed sub-lattice interactions of a single atom in one layer with all relevant atoms in the other layer of the bilayer systems. While graphite and *h*-BN present different optimal stacking modes the vdW energy only weakly depends on the exact staking of the hexagonal lattices.[17] Therefore, since the C-C and B-N $C_6$ coefficients were found to be very similar, the vdW contributions of the mixed interactions of both systems are nearly identical. Similarly, panel (b) of Fig. 5 shows the vdW energy contribution of the same sub-lattice interactions. Here, due to the large differences between the C-C, B-B, and N-N $C_6$ coefficients the separate contribution of each of the sub-lattice interactions is quite different. Nevertheless, when adding the contributions of the BB and NN interactions in *h*-BN and the CC and C'C' interactions in bilayer graphene the overall contributions are very similar reflecting the fact that the C-C $C_6$ coefficient is not far from the average value of the B-B and N-N coefficients. Thus, as shown in panel (c) of Fig. 3, owing to the isoelectronic nature of the two materials, their similar intra-layer bond lengths and lattice structures, and the ordering of the atomic static polarizabilities, the overall vdW attraction per atom in the unit-cell of bilayer graphene and *h*-BN are very similar despite the differences in the individual $C_6$ coefficients of the different atoms.

Finally, these results are clearly manifested in the full binding energy curves presented in panel (d) of Fig. 5 for bulk graphite (calculated by F. Hanke[8]) and *h*-BN (calculated by Marom *et al.*[7]) as obtained using the TS-vdW scheme. As can be seen, both binding energy curves predict the same interlayer distance of 3.33 Å in excellent agreement with the experimental values[45-53] and similar binding energies (graphite: 84.7 meV/atom; *h*-BN: 85.9 meV/atom). The dispersive attractive part of both systems is very similar whereas the main deviations between the two curves appear in the short-range where Pauli-repulsions due to overlap of the B-N electron clouds in *h*-BN and C-C electron clouds in graphite become dominant. These deviations are to be expected as the two materials possess different optimal stacking modes and since the effective volumes of carbon in graphite and boron and nitrogen in *h*-BN are different.

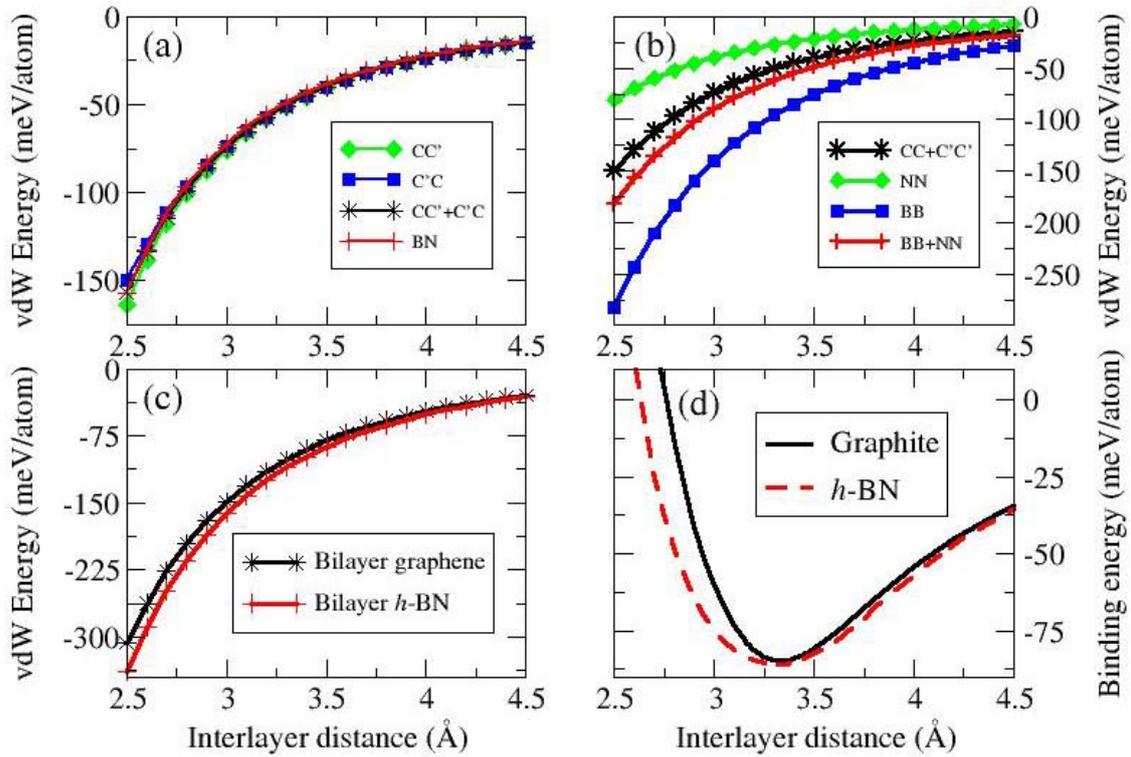

Figure 5: vdW contributions to the binding energy curves of graphite and *h*-BN in the AB and AA' stacking modes, respectively. (a) Mixed sub-lattice contributions to the vdW energy of the bilayer systems. CC' represent the interaction of a single carbon atom located on sub-lattice 1 of the first layer with all carbon atoms located on sub-lattice 2 of the second layer. C'C represents the interaction of a single carbon atom located on sub-lattice 2 of the first layer with all carbon atoms located on sub-lattice 1 of the second layer. CC'+C'C represents the overall sum of the CC' and C'C contribution per atom. BN stands for the interaction of a single boron(nitrogen) atom in one *h*-BN layer with all nitrogen(boron) atoms in the other layer. (b) Same sub-lattice contributions to the vdW energy of the bilayer systems. CC+C'C' represent the overall sum of the interaction of a single carbon atom located on sub-lattice 1 of the first layer with all carbon atoms located on the same sub-lattice of the second layer and the interaction of a single carbon atom located on sub-lattice 2 of the first layer with all carbon atoms located on sub-lattice 2 of the second layer. Due to the symmetry of the hexagonal lattice the CC and C'C' contributions are identical to each other and therefore also to the CC+C'C' contribution per atom. NN(BB) represents the interaction of a single boron(nitrogen) atom in one layer with all boron(nitrogen) atoms in the second layer. BB+NN represents the sum of the BB and NN vdW contributions per atom. (c) Total vdW energy per atom of the bilayer systems. See supporting material for further details regarding this calculation. (d) Full binding energy curves of bulk graphite[8] (solid black line) and bulk *h*-BN[7] (red dashed line) as calculated using the TS-vdW method. Results for the graphite binding energy calculations have been provided courtesy of Felix Hanke.

To summarize, in the present study the interlayer binding in graphene and *h*-BN were compared. It was found that despite the polar nature of the B-N bond in *h*-BN, full lattice sum of the electrostatic contributions from the formal charges on all atomic sites within the layer results in rapid exponential decay of the electrostatic potential into the vacuum. As a result, at the equilibrium interlayer distance the overall monopolar electrostatic contribution to the interlayer binding is a small fraction of the total calculated binding energy. At physically relevant interlayer distances, the contribution of higher order electrostatic multipoles and exchange-SR-correlation energies elegantly cancels out with the kinetic energy term manifesting the effect of Pauli repulsions. Nonetheless, despite the effective marginal contribution of electrostatic interactions to the interlayer binding, when considering relative lateral shifts of the layers at the equilibrium interlayer distance, the residual electrostatic potential along with the Pauli repulsions are sufficient to set the AA' stacking mode as the optimal configuration of *h*-BN. The opposite holds true for the dispersive component which has a minor effect on the corrugation of the interlayer sliding energy surface[7,17] but is a crucial factor for the interlayer anchoring in both systems.[7,16] Here, despite notable differences between the free-atom as well as the bulk homo-nuclear $C_6$ coefficients of the two materials, the hetero-atomic bulk coefficients in *h*-BN agree to within 2% with the C-C coefficients in bulk graphite. This translates to very similar binding energy curves for both materials (deviating mainly at distances shorter than the equilibrium interlayer distance where Pauli repulsions become dominant) thus resulting in similar binding energies and practically identical equilibrium interlayer distances for graphene and *h*-BN. These conclusions are further supported by recent studies of *h*-BN/graphene hybrid structures[91-97] which, similar to graphite and h-BN, are predicted to present an interlayer distance of 3.3 Å.[92]

Some notes regarding the calculations presented in this study should be made: (i) when performing the electrostatic (and vdW) sums only the pristine systems have been considered. Defects, such as lattice vacancies,[33,98] may introduce long range effective Coulomb potentials which decay asymptotically as 1/r rather than exponentially into the vacuum; and (ii) the bulk TS-vdW calculations presented above lack a proper description of the screening of the pair-wise interaction by the dielectric medium and non-additive many-body vdW energy contributions. The neglect of screening effects usually results in too large bulk $C_6$ coefficients and therefore overestimated binding energy values but often gives accurate predictions for interlayer distances.[7-8,15,23-24,99-102] As screening effects on the unscreened $C_6$ coefficients are expected to be similar in graphite and *h*-BN, which have the same intra-layer hexagonal lattice structure, the inclusion of such effects is expected to influence the binding energy curves of both materials in a similar manner thus leaving the conclusions drawn here, based on the unscreened coefficients, valid. The proper description of all the above mentioned effects is a subject of ongoing research.

Finally, a note should be made regarding the general nature of the conclusions drawn above. The rapid decay of the monopolar electrostatic potential into the vacuum above the two dimensional layer is not a unique property of the hexagonal lattice of *h*-BN.[62] While its fine details are expected to depend on the chemical composition and structural topology of the underlying material, the general nature of the exponential decay is expected to prevail in many layered systems (see supporting material). In contrast, the contribution of higher-order electrostatic multipoles, exchange-SR-correlation energies, and Pauli repulsions at different interlayer distances may heavily depend on the specific chemical nature of the material and its lattice structure. Therefore, when studying the interlayer binding in such materials, the careful balance between electrostatic, dispersive, and Pauli interactions should be considered.

# Acknowledgments

The author would like to thank Dr. Alexandre Tkatchenko and Dr. Felix Hanke for generously sharing results of their calculations and for many valuable discussions. Many thanks to Prof. Ernesto Joselevich, Prof. Leeor Kronik, Prof. Haim Diamant, Prof. Shahar Hod, Prof. Fernando Patolsky, Dr. Noa Marom, Mr. Guy Cohen, and Mr. Tal Levy for fruitful discussions regarding the subject. This work was supported by the Israel Science Foundation under grant No. 1313/08, the Center for Nanoscience and Nanotechnology at Tel Aviv University, and the Lise Meitner-Minerva Center for Computational Quantum Chemistry. The research leading to these results has received funding from the European Community's Seventh Framework Programme FP7/2007-2013 under grant agreement No. 249225.

# Convergence tests for the cluster DFT calculations

1. <u>Convergence with respect to basis set.</u>

Test calculations for basis set convergence have been performed using the PBE functional for a 72 atoms zigzag *h*-BN dimmer. A set of three Gaussian basis sets with increasing size and diffuseness has been used including the 3-21G, 6-31G**, and 6-311++G(3df,3pd). At an interlayer distance of 3.3 Å the difference between the total energy calculated using the 6-31G** and the 6-311++G(3df,3pd) basis sets is ~1.3 meV/atom. When looking at the different energy components at that interlayer distance the exchange-correlation energy is converged down to ~1.2 meV/atom. The electrostatic and kinetic energies are less converged (~8 and ~10.6 meV/atom, respectively) although, as mentioned above, their overall contribution to the total energy is well converged. The following figure shows the convergence of the different energy terms as function of interlayer distance with respect to the basis set used. As can be seen, for the purpose of estimating the role of the different energy components for the interlayer binding the 6-31G** basis set results are satisfying.

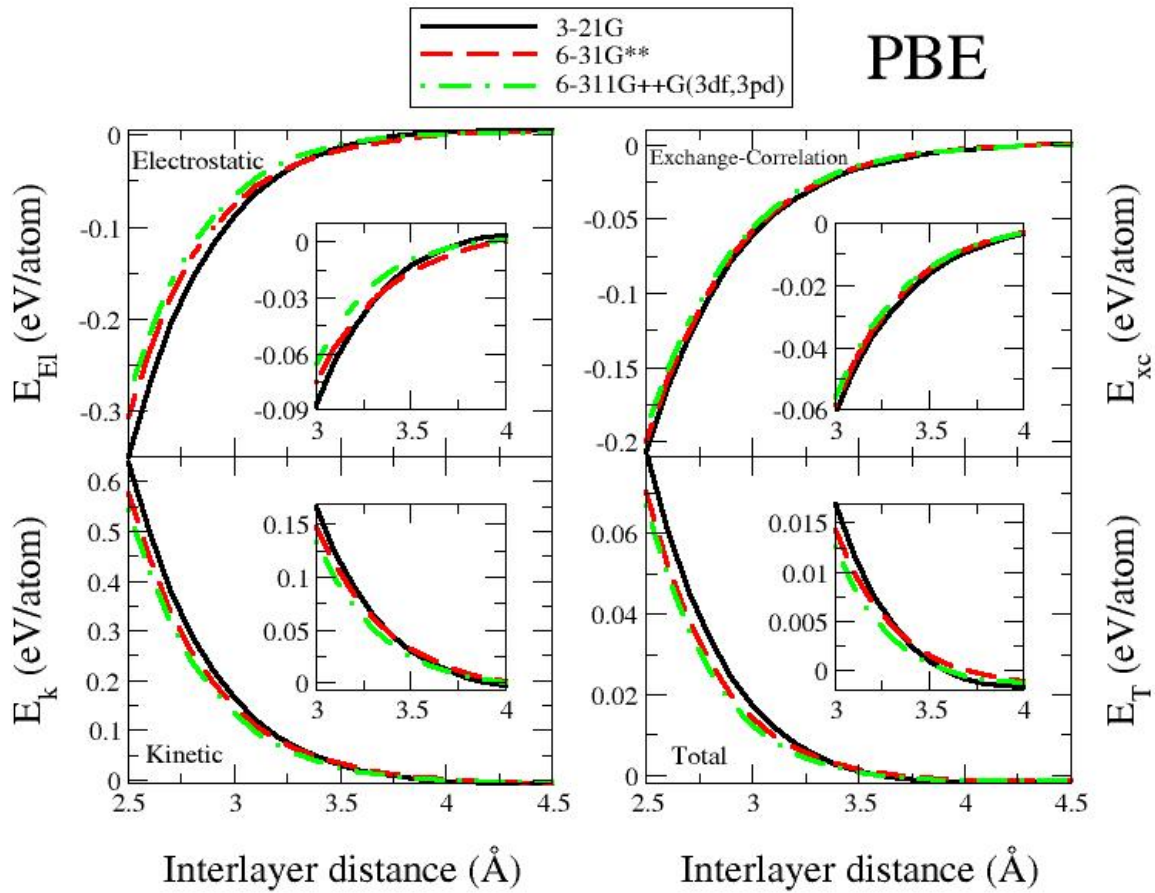

Fig. S1: Basis set convergence tests for the cluster DFT calculations. Electrostatic (upper left panel), exchange correlation (upper right panel), kinetic (lower left panel), and total (lower right panel) energies as a function of the interlayer distance of a 72 atoms zigzag *h*-BN dimmer calculated using the PBE functional and the 3-21G (solid black line), 6-31G** (dashed red line), and the 6-311++G(3df,3pd) (dashed-dotted green line) basis sets. Insets: zoom in on the region of physically relevant interlayer distances.

2. Convergence with respect to the size and shape of the cluster

To test for convergence of the results with respect to the shape and size of the finite bilayer flakes used, calculations of the different energy components as a function of the interlayer distance for armchair and zigzag flakes of increasing size have been performed. Fig. S2 presents images of some of the flakes used in the present study.

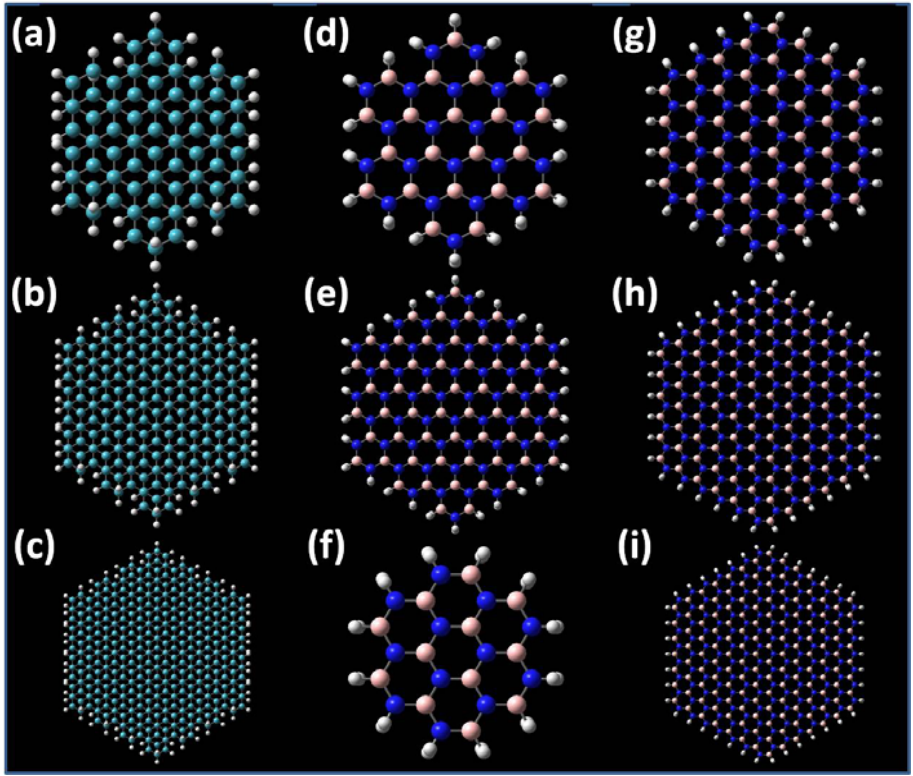

Fig. S2: Images of some of the hydrogen terminated dimmer hexagonal flakes used in the present study: (a) 120 atoms armchair graphene dimmer, (b) 288 atoms armchair graphene dimmer, (c) 528 atoms armchair graphene dimmer, (d) 120 atoms armchair $h$-BN dimmer, (e) 288 atoms armchair $h$-BN dimmer, (f) 72 atoms zigzag $h$-BN dimmer, (g) 240 atoms zigzag $h$-BN dimmer, (h) 504 atoms zigzag $h$-BN dimmer, and (i) 672 atoms zigzag $h$-BN dimmer. Cyan, blue, pink and grey spheres represent carbon, nitrogen, boron, and hydrogen atoms, respectively.

All calculations have been performed using the PBE/6-31G** level of theory. At 3.3 Å the total, electrostatic, exchange-correlation, and kinetic energies of the zigzag $h$-BN dimmer consisting of 672 atoms are converged down to 0.2, 1.4, 0.5, 2.1 meV/atom, respectively. For the armchair graphene dimmer consisting of 528 atoms at the same interlayer distance the total, electrostatic, exchange-correlation, and kinetic energies are converged down to 0.2, 2.5, 1.7, 4.4 meV/atom, respectively.

Figs. S3 and S4 present the convergence of the interlayer dependence of the different energy components as a function of flake size for bilayer graphene and $h$-BN, respectively. As can be seen, for all practical purposes the results of the largest cluster sizes are well converged showing only marginal edge effects.

Fig. S5 compares the interlayer dependence of the different energy components of two armchair and one zigzag $h$-BN dimmers showing that the converged results are independent of the shape of the flakes used in the calculations.

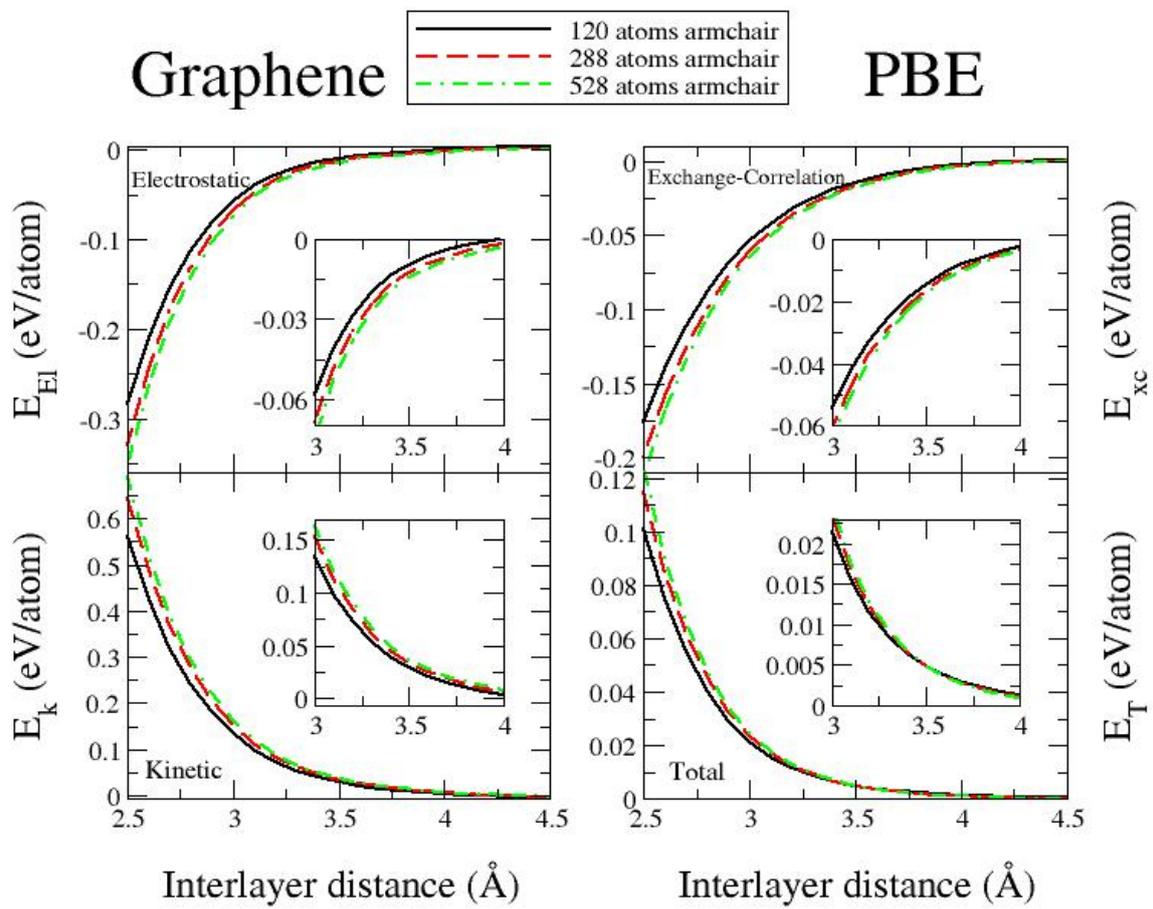

Fig. S3: Convergence of the interlayer dependence of the electrostatic (upper left), exchange-correlation (upper right), kinetic (lower left), and total (lower right) energies with respect to the size of the armchair graphene flakes. Solid black, dashed red, and dashed-dashed-dotted green curves represent results for the 120, 288, and 528 atoms graphene dimmer flakes.

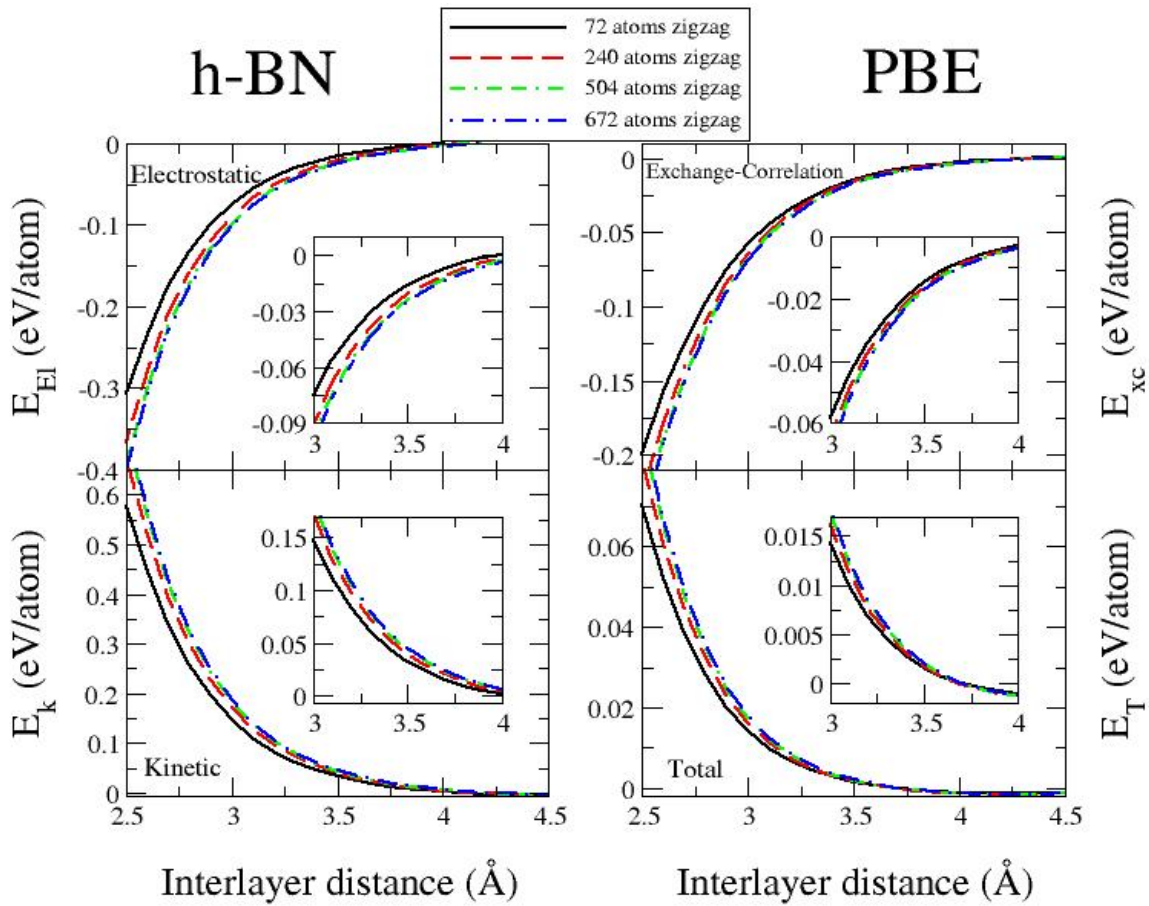

Fig. S4: Convergence of the interlayer dependence of the electrostatic (upper left), exchange-correlation (upper right), kinetic (lower left), and total (lower right) energies with respect to the size of the zigzag $h$-BN flakes. Solid black, dashed red, dashed-dashed-dotted green, and dashed-dotted blue curves represent results for the 72, 240, 504, and 672 atoms $h$-BN dimmer flakes.

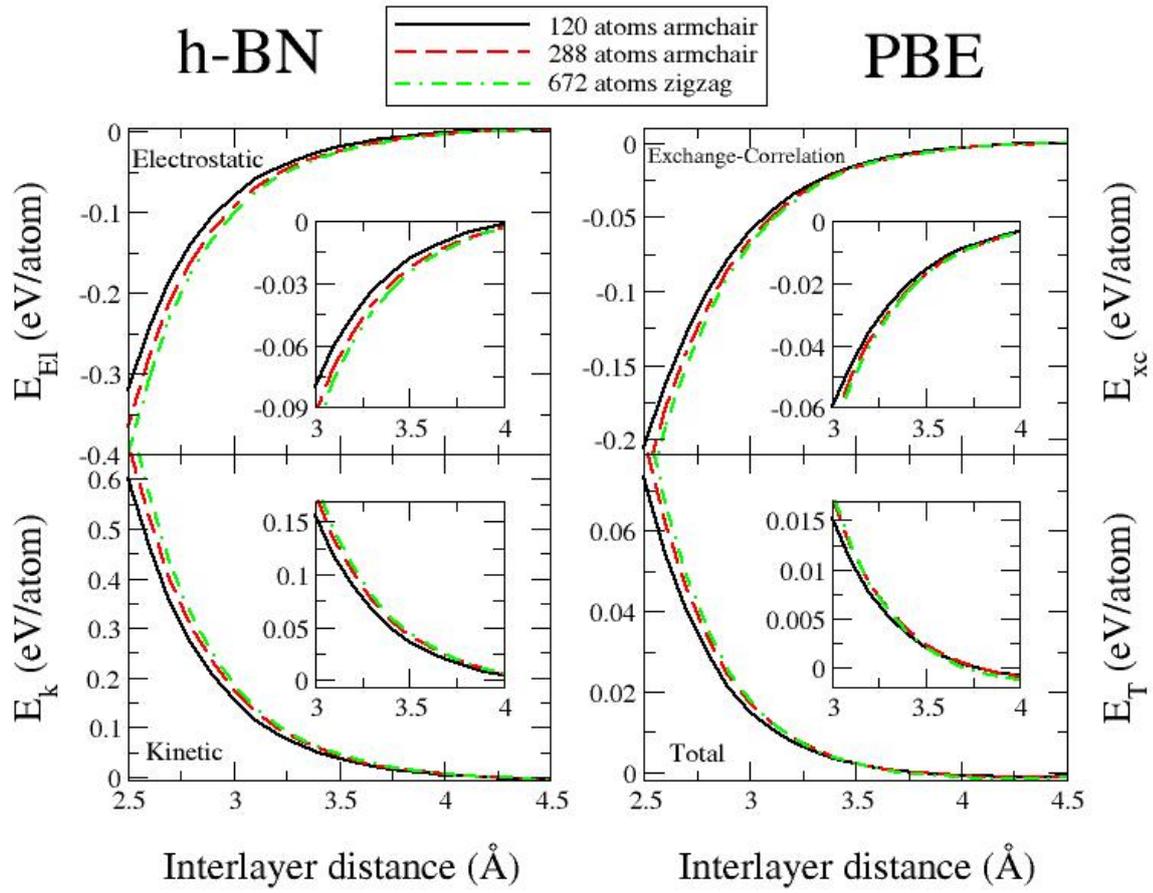

Fig. S5: Convergence of the interlayer dependence of the electrostatic (upper left), exchange-correlation (upper right), kinetic (lower left), and total (lower right) energies with respect to the shape of the *h*-BN flakes. Solid black, dashed red, and dashed-dashed-dotted green curves represent results for the 120 atoms armchair, 288 atoms armchair, and 672 atoms zigzag *h*-BN dimmer flakes.

# The Ewald summation method for the Coulomb potential in two dimensional (2D) periodic slab geometries

We are interested in calculating the electrostatic potential induced by a slab of charged particles which is periodic in two directions and finite in the third direction. For simplicity we consider a rectangular unit-cell such that the periodic directions are aligned along the X and Y axes with translational vectors $\vec{a}_x = (T_x, 0, 0)$ and $\vec{a}_y = (0, T_y, 0)$. The unit cell is assumed to contain $d$ charged particles located at positions $\vec{r}_i = (x_i, y_i, z_i)$; $i = 1, 2, \ldots, d$ and to be charge neutral such that:

(1) $\sum_{i=1}^{d} q_i = 0$

The location of a general atom in the slab is then given by the following expression:

(2) $\vec{r}_{i,n,m} = \vec{r}_i + n\vec{a}_x + m\vec{a}_y = (x_i + nT_x, y_i + mT_y, z_i)$; $i = 1, 2, \ldots, d$; $n_1, n_2 = 0, \pm 1, \pm 2, \cdots, \pm \infty$

With these definitions we may write the general expression for the electrostatic potential at point $\vec{r} = (x, y, z) \neq \vec{r}_{i,n,m}$ due to the infinite 2D-periodic slab as:

(3) $\varphi(\vec{r}) = \varphi(x, y, z) = \sum_{i=1}^{d} \sum_{n=-\infty}^{\infty} \sum_{m=-\infty}^{\infty} \frac{q_i}{|\vec{r} - \vec{r}_{i,n,m}|} = \sum_{i=1}^{d} \sum_{n=-\infty}^{\infty} \sum_{m=-\infty}^{\infty} \frac{q_i}{\sqrt{(x - x_i - nT_x)^2 + (y - y_i - mT_y)^2 + (z - z_i)^2}}$

The potential $\varphi(\vec{r})$ is a periodic function in the X and Y directions with a period of $T_x = |\vec{a}_x|$ along the X direction and a period of $T_y = |\vec{a}_y|$ along the Y direction. This can be easily demonstrated as follows:

(4) $\varphi(x + pT_x, y, z) = \sum_{i=1}^{d} \sum_{n=-\infty}^{\infty} \sum_{m=-\infty}^{\infty} \frac{q_i}{[(x + pT_x - x_i - nT_x)^2 + (y - y_i - mT_y)^2 + (z - z_i)^2]^{1/2}} =$

$= \sum_{i=1}^{d} \sum_{n=-\infty}^{\infty} \sum_{m=-\infty}^{\infty} \frac{q_i}{[(x - x_i - (n-p)T_x)^2 + (y - y_i - mT_y)^2 + (z - z_i)^2]^{1/2}} \overset{l=n-p}{=}$

$= \sum_{i=1}^{d} \sum_{l=-\infty}^{\infty} \sum_{m=-\infty}^{\infty} \frac{q_i}{[(x - x_i - lT_x)^2 + (y - y_i - mT_y)^2 + (z - z_i)^2]^{1/2}} = \varphi(x, y, z)$

where $p$ is an integer number and the equality leading from the second to the third line uses the fact the summation over the index n covers the infinite range of integer numbers from $-\infty$ to $+\infty$. Similar arguments may be used to prove the periodicity in the y direction.

We can now define the potential arising at point $\vec{r}$ due to one of the particles in the unit cell and its periodic images:

(5)
$$\varphi_i(\vec{r}) = \varphi_i(x,y,z) = \sum_{n=-\infty}^{\infty}\sum_{m=-\infty}^{\infty} \frac{q_i}{\sqrt{(x-x_i-nT_x)^2+(y-y_i-mT_y)^2+(z-z_i)^2}} =$$

$$= q_i \sum_{n=-\infty}^{\infty}\sum_{m=-\infty}^{\infty} \frac{1}{\sqrt{(x-x_i-nT_x)^2+(y-y_i-mT_y)^2+(z-z_i)^2}}$$

such that

(6) $\varphi(\vec{r}) = \sum_{i=1}^{d} \varphi_i(\vec{r})$

The sum appearing in Eq. (5) is divergent while the sum appearing in Eq. (3) is conditionally convergent provided that the unit cell is charge neutral (Eq. (1)). In order to evaluate the sum $\varphi(\vec{r})$ we utilize the Ewald summation method which splits the conditionally convergent sum into two absolutely (fast) converging sums. This is achieved by splitting the singular and long-range Coulomb potential into a singular short-range part, which can be readily evaluated in real-space, and a non-singular long-range part which is evaluated in reciprocal space. To this end, we notice that, similar to $\varphi(\vec{r})$, $\varphi_i(\vec{r})$ is also a periodic function in the X and Y directions with the same periodicity (the proof follow the same lines as the proof for $\varphi(\vec{r})$).

We now recall that a periodic function $f(x) = f(x+nT); n = 0,\pm 1,\pm 2,\ldots$ can be expanded in a Fourier series of the form:

(7) $f(x) = \sum_{m=-\infty}^{\infty} \hat{f}(m) e^{2\pi i m x/T}$ ; $\hat{f}(m) = \frac{1}{T}\int_{-T/2}^{T/2} f(x) e^{-2\pi i m x/T} dx$

Defining $k_x = 0, \pm 2\pi/T, \pm 4\pi/T, \ldots$ the expansion can be rewritten in the following form:

(8) $f(x) = \sum_{k_x} \hat{f}(k_x) e^{ik_x x}$ ; $\hat{f}(k_x) = \frac{1}{T}\int_{-T/2}^{T/2} f(x) e^{-ik_x x} dx$ ; $k_x = 0, \pm 2\pi/T, \pm 4\pi/T, \ldots$

Similarly, for a two-dimensional (2D) periodic function which obeys $f(x,y) = f(x+nT_x, y+mT_y)$ ; $n,m = 0,\pm 1,\pm 2,\ldots$ one has:

(9)
$$f(x,y) = \sum_{k_x}\sum_{k_y} \hat{f}(k_x,k_y) e^{i(k_x x + k_y y)} = \sum_{k_x}\sum_{k_y} \hat{f}(\vec{\kappa}) e^{i\vec{\kappa}\cdot\vec{\rho}}$$

$$\hat{f}(\vec{\kappa}) = \frac{1}{T_x}\frac{1}{T_y}\int_{-T_x/2}^{T_x/2} dx \int_{-T_y/2}^{T_y/2} dy\, f(x,y) e^{-i\vec{\kappa}\cdot\vec{\rho}}$$

With

(10) $\vec{\rho} = (x,y); \vec{\kappa} = (k_x, k_y)$
$k_x = 0, \pm 2\pi/T_x, \pm 4\pi/T_x, \ldots$
$k_y = 0, \pm 2\pi/T_y, \pm 4\pi/T_y, \ldots$

It is now possible to expand $\varphi_i(\vec{r})$ as a 2D Fourier series of the form:

(11) $$\varphi_i(\vec{r}) = \sum_{k_x}\sum_{k_y} \hat{\varphi}_i(\vec{\kappa},z) e^{i\vec{\kappa}\cdot\vec{\rho}}$$

With the Fourier coefficients given by:

(12) $$\hat{\varphi}_i(\vec{\kappa},z) = \frac{1}{T_x T_y} \int_{-T_x/2}^{T_x/2} dx \int_{-T_y/2}^{T_y/2} dy\, \varphi_i(\vec{r}) e^{-i\vec{\kappa}\cdot\vec{\rho}} =$$
$$= \frac{q_i}{T_x T_y} \sum_{n=-\infty}^{\infty}\sum_{m=-\infty}^{\infty} \int_{-T_x/2}^{T_x/2} dx \int_{-T_y/2}^{T_y/2} dy \frac{1}{\sqrt{(x-x_i-nT_x)^2 + (y-y_i-mT_y)^2 + (z-z_i)^2}} e^{-i(k_x x + k_y y)}$$

Defining new variables:

(13) $$\begin{cases} \tilde{x} = x - x_i - nT_x \\ \tilde{y} = y - y_i - mT_y \\ \tilde{z} = z - z_i \end{cases} \Rightarrow \begin{cases} d\tilde{x} = dx \\ d\tilde{y} = dy \\ d\tilde{z} = dz \end{cases} ; \begin{cases} x = \tilde{x} + x_i + nT_x \\ y = \tilde{y} + y_i + mT_y \\ z = \tilde{z} + z_i \end{cases}$$

We can rewrite the integral as:

(14) $$\hat{\varphi}_i(\vec{\kappa},\tilde{z}) = \frac{q_i}{T_x T_y} \sum_{n=-\infty}^{\infty}\sum_{m=-\infty}^{\infty} \int_{-T_x/2-x_i-nT_x}^{T_x/2-x_i-nT_x} d\tilde{x} \int_{-T_y/2-y_i-mT_y}^{T_y/2-y_i-mT_y} d\tilde{y} \frac{1}{\sqrt{\tilde{x}^2+\tilde{y}^2+\tilde{z}^2}} e^{-i[k_x(\tilde{x}+x_i+nT_x)+k_y(\tilde{y}+y_i+mT_y)]} =$$
$$= \frac{q_i}{T_x T_y} e^{-i(k_x x_i + k_y y_i)} \sum_{n=-\infty}^{\infty}\sum_{m=-\infty}^{\infty} \int_{-T_x/2-x_i-nT_x}^{T_x/2-x_i-nT_x} d\tilde{x} \int_{-T_y/2-y_i-mT_y}^{T_y/2-y_i-mT_y} d\tilde{y} \frac{1}{\sqrt{\tilde{x}^2+\tilde{y}^2+\tilde{z}^2}} e^{-i(k_x \tilde{x}+k_y \tilde{y})} e^{-i(k_x nT_x + k_y mT_y)}$$

Noticing that $e^{-i(k_x nT_x + k_y mT_y)} = e^{-i\left(\frac{2\pi p}{T_x}nT_x + \frac{2\pi q}{T_y}mT_y\right)} = e^{-2\pi i(pn+qm)} = 1$; $p,q,n,m = 0,\pm 1,\pm 2,\ldots$ and defining $\vec{\rho}_i = (x_i, y_i)$ and $\vec{\tilde{\rho}} = (\tilde{x}, \tilde{y})$ we obtain:

(15) $$\hat{\varphi}_i(\vec{\kappa},\tilde{z}) = \frac{q_i}{T_x T_y} e^{-i\vec{\kappa}\cdot\vec{\rho}_i} \sum_{n=-\infty}^{\infty}\sum_{m=-\infty}^{\infty} \int_{-T_x/2-x_i-nT_x}^{T_x/2-x_i-nT_x} d\tilde{x} \int_{-T_y/2-y_i-mT_y}^{T_y/2-y_i-mT_y} d\tilde{y} \frac{1}{\sqrt{\tilde{x}^2+\tilde{y}^2+\tilde{z}^2}} e^{-i\vec{\kappa}\cdot\vec{\tilde{\rho}}} =$$
$$= \frac{q_i}{T_x T_y} e^{-i\vec{\kappa}\cdot\vec{\rho}_i} \sum_{n=-\infty}^{\infty} \int_{-T_x/2-x_i-nT_x}^{T_x/2-x_i-nT_x} d\tilde{x} \sum_{m=-\infty}^{\infty} \int_{-T_y/2-y_i-mT_y}^{T_y/2-y_i-mT_y} d\tilde{y} \frac{1}{\sqrt{\tilde{x}^2+\tilde{y}^2+\tilde{z}^2}} e^{-i\vec{\kappa}\cdot\vec{\tilde{\rho}}}$$

In Eq. (15) both integrations are performed over finite segments. The summation shifts the segments in a consecutive manner such that the sum of all segmental integrals can be replaced by an integral over the full range:

(16) $$\sum_{n=-\infty}^{\infty} \int_{-T_x/2-x_i-nT_x}^{T_x/2-x_i-nT_x} d\tilde{x} = \cdots + \int_{-3T_x/2-x_i}^{-T_x/2-x_i} d\tilde{x} + \int_{-T_x/2-x_i}^{T_x/2-x_i} d\tilde{x} + \int_{T_x/2-x_i}^{3T_x/2-x_i} d\tilde{x} + \cdots = \int_{-\infty}^{\infty} d\tilde{x}$$

Here, we have displayed the terms with $n=1,0,-1$, respectively. Similarly, we obtain $\sum_{m=-\infty}^{\infty}\int_{-T_y/2-y_i-mT_y}^{T_y/2-y_i-mT_y} d\tilde{y} = \int_{-\infty}^{\infty} d\tilde{y}$ such that $\hat{\varphi}(\vec{\kappa},\tilde{z})$ is given by:

$$(17) \quad \hat{\varphi}_i(\vec{\kappa},\tilde{z}) = \frac{q_i}{T_x T_y} e^{-i\vec{\kappa}\cdot\vec{\rho}_i} \int_{-\infty}^{\infty} d\tilde{x} \int_{-\infty}^{\infty} d\tilde{y} \frac{1}{\sqrt{\tilde{x}^2 + \tilde{y}^2 + \tilde{z}^2}} e^{-i\vec{\kappa}\cdot\vec{\rho}}$$

The evaluation of the remaining double integrals is performed by transforming to a polar coordinate system such that:

$$(18) \quad \begin{cases} \tilde{\rho} = |\vec{\tilde{\rho}}| = \sqrt{\tilde{x}^2 + \tilde{y}^2} \\ \tilde{\theta} = arctg\left(\frac{\tilde{y}}{\tilde{x}}\right) \end{cases} ; \quad \begin{cases} \tilde{x} = \tilde{\rho}\cos(\tilde{\theta}) \\ \tilde{y} = \tilde{\rho}\sin(\tilde{\theta}) \end{cases}$$

and

$$(19) \quad \hat{\varphi}_i(\vec{\kappa},\tilde{z}) = \frac{q_i}{T_x T_y} e^{-i\vec{\kappa}\cdot\vec{\rho}_i} \int_0^{\infty} \frac{\tilde{\rho}}{\sqrt{\tilde{\rho}^2 + \tilde{z}^2}} d\tilde{\rho} \int_0^{2\pi} d\tilde{\theta} e^{-i(k_x \tilde{\rho}\cos\tilde{\theta} + k_y \tilde{\rho}\sin\tilde{\theta})} =$$
$$= \frac{q_i}{T_x T_y} e^{-i\vec{\kappa}\cdot\vec{\rho}_i} \int_0^{\infty} \frac{\tilde{\rho}}{\sqrt{\tilde{\rho}^2 + \tilde{z}^2}} d\tilde{\rho} \int_0^{2\pi} d\tilde{\theta} e^{-i(k_x \cos\tilde{\theta} + k_y \sin\tilde{\theta})\tilde{\rho}}$$

We may now also transform $\vec{\kappa}$ to its polar representation

$$(20) \quad \begin{cases} \kappa = |\vec{\kappa}| = \sqrt{k_x^2 + k_y^2} \\ \eta = arctg\left(\frac{k_y}{k_x}\right) \end{cases} ; \quad \begin{cases} k_x = \kappa\cos(\eta) \\ k_y = \kappa\sin(\eta) \end{cases}$$

to obtain:

$$(21) \quad \hat{\varphi}_i(\vec{\kappa},\tilde{z}) = \frac{q_i}{T_x T_y} e^{-i\vec{\kappa}\cdot\vec{\rho}_i} \int_0^{\infty} \frac{\tilde{\rho}}{\sqrt{\tilde{\rho}^2 + \tilde{z}^2}} d\tilde{\rho} \int_0^{2\pi} d\tilde{\theta} e^{-i(\cos\eta\cos\tilde{\theta} + \sin\eta\sin\tilde{\theta})\kappa\tilde{\rho}} =$$
$$= \frac{q_i}{T_x T_y} e^{-i\vec{\kappa}\cdot\vec{\rho}_i} \int_0^{\infty} \frac{\tilde{\rho}}{\sqrt{\tilde{\rho}^2 + \tilde{z}^2}} d\tilde{\rho} \int_0^{2\pi} d\tilde{\theta} e^{-i\kappa\tilde{\rho}\cos(\tilde{\theta}-\eta)}$$

By changing variables such that $\alpha = \tilde{\theta} - \eta$; $d\alpha = d\tilde{\theta}$ we can write:

$$(22) \quad \hat{\varphi}_i(\vec{\kappa},\tilde{z}) = \frac{q_i}{T_x T_y} e^{-i\vec{\kappa}\cdot\vec{\rho}_i} \int_0^{\infty} \frac{\tilde{\rho}}{[\tilde{\rho}^2 + \tilde{z}^2]^{1/2}} d\tilde{\rho} \int_{-\eta}^{2\pi-\eta} e^{-i\kappa\tilde{\rho}\cos\alpha} d\alpha$$

Since the cosine function is periodic and the angular integration is performed over a full period of $2\pi$ we may write:

$$\text{(23)} \quad \int_{-\eta}^{2\pi-\eta} e^{-i\kappa\tilde{\rho}\cos\alpha}\,d\alpha = \int_{0}^{2\pi} e^{-i\kappa\tilde{\rho}\cos\alpha}\,d\alpha = 2\pi J_0(|\kappa\tilde{\rho}|) = 2\pi J_0(\kappa\tilde{\rho})$$

where $J_0(x)$ is the zeroth order Bessel function of the first kind, and the last equality results from the fact that $\kappa$ and $\tilde{\rho}$ are positive norms of the corresponding vectors. With this we can write:

$$\text{(24)} \quad \hat{\varphi}_i(\vec{\kappa},\tilde{z}) = \frac{2\pi q_i}{T_x T_y} e^{-i\vec{\kappa}\cdot\vec{\rho}_i} \int_0^{\infty} \frac{\tilde{\rho} J_0(\kappa\tilde{\rho})}{\sqrt{\tilde{\rho}^2 + \tilde{z}^2}}\,d\tilde{\rho} = \frac{2\pi q_i}{T_x T_y} e^{-i\vec{\kappa}\cdot\vec{\rho}_i} \int_0^{\infty} \frac{(R/\kappa) J_0(R)}{\left[R^2/\kappa^2 + \tilde{z}^2\right]^{1/2}} \frac{dR}{\kappa} =$$
$$= \frac{2\pi q_i}{T_x T_y} \frac{e^{-i\vec{\kappa}\cdot\vec{\rho}_i}}{\kappa} \int_0^{\infty} \frac{R J_0(R)}{\left[R^2 + (\tilde{z}\kappa)^2\right]^{1/2}}\,dR = \frac{2\pi q_i}{T_x T_y} \frac{e^{-i\vec{\kappa}\cdot\vec{\rho}_i}}{\kappa} e^{-\kappa|\tilde{z}|}$$

where we have set $R = \kappa\tilde{\rho}$; $dR = \kappa d\tilde{\rho}$. Since, $\tilde{z} = z - z_i$ we can finally write:

$$\text{(25)} \quad \hat{\varphi}_i(\vec{\kappa},z) = \frac{2\pi q_i}{T_x T_y} e^{-i\vec{\kappa}\cdot\vec{\rho}_i} \frac{e^{-\kappa|z-z_i|}}{\kappa}$$

In order to construct fully convergent lattice sums, we now use the following integral identity to split the $1/\kappa$ into two ranges:

$$\text{(26)} \quad \frac{e^{-\kappa|z|}}{\kappa} = \frac{2}{\sqrt{\pi}} \int_0^{\infty} e^{-\kappa^2 t^2 - |z-z_i|^2/(4t^2)}\,dt = \frac{2}{\sqrt{\pi}} \int_0^{\lambda} e^{-\kappa^2 t^2 - |z-z_i|^2/(4t^2)}\,dt + \frac{2}{\sqrt{\pi}} \int_{\lambda}^{\infty} e^{-\kappa^2 t^2 - |z-z_i|^2/(4t^2)}\,dt$$

Using this we obtain:

$$\text{(27)} \quad \hat{\varphi}_i(\vec{\kappa},z) = \frac{2\pi q_i}{T_x T_y} e^{-i\vec{\kappa}\cdot\vec{\rho}_i} \left[\frac{2}{\sqrt{\pi}} \int_0^{\lambda} e^{-\kappa^2 t^2 - |z-z_i|^2/(4t^2)}\,dt + \frac{2}{\sqrt{\pi}} \int_{\lambda}^{\infty} e^{-\kappa^2 t^2 - |z-z_i|^2/(4t^2)}\,dt\right] =$$
$$= \frac{4\sqrt{\pi} q_i}{T_x T_y} e^{-i\vec{\kappa}\cdot\vec{\rho}_i} \int_0^{\lambda} e^{-\kappa^2 t^2 - |z-z_i|^2/(4t^2)}\,dt + \frac{4\sqrt{\pi} q_i}{T_x T_y} e^{-i\vec{\kappa}\cdot\vec{\rho}_i} \int_{\lambda}^{\infty} e^{-\kappa^2 t^2 - |z-z_i|^2/(4t^2)}\,dt = \hat{\varphi}_i^S(\vec{\kappa},z) + \hat{\varphi}_i^L(\vec{\kappa},z)$$

Where the short (S) and long (L) range contributions to $\hat{\varphi}_i(\vec{\kappa},z)$ have been defined as:

$$\text{(28)} \quad \begin{cases} \hat{\varphi}_i^S(\vec{\kappa},z) = \dfrac{4\sqrt{\pi} q_i}{T_x T_y} e^{-i\vec{\kappa}\cdot\vec{\rho}_i} \displaystyle\int_0^{\lambda} e^{-\kappa^2 t^2 - |z-z_i|^2/(4t^2)}\,dt \\ \hat{\varphi}_i^L(\vec{\kappa},z) = \dfrac{4\sqrt{\pi} q_i}{T_x T_y} e^{-i\vec{\kappa}\cdot\vec{\rho}_i} \displaystyle\int_{\lambda}^{\infty} e^{-\kappa^2 t^2 - |z-z_i|^2/(4t^2)}\,dt \end{cases}$$

The full potential can be now written as:

$$\text{(29)} \quad \varphi(\vec{\kappa},z) = \sum_{i=1}^{d} \varphi_i(\vec{\kappa},z) = \sum_{i=1}^{d} \left[\hat{\varphi}_i^S(\vec{\kappa},z) + \hat{\varphi}_i^L(\vec{\kappa},z)\right] = \sum_{i=1}^{d} \hat{\varphi}_i^S(\vec{\kappa},z) + \sum_{i=1}^{d} \hat{\varphi}_i^L(\vec{\kappa},z) = \varphi^S(\vec{\kappa},z) + \varphi^L(\vec{\kappa},z)$$

## The long range term

The long range term, $\varphi^L(\vec{r})$, is absolutely convergent in reciprocal space and thus it is first summed in reciprocal space and then transformed back to real space. When evaluating the relevant integrals one needs to separately treat the cases where $\kappa > 0$ and $\kappa = 0$:

### $\kappa > 0$

For $\kappa > 0$ the long-range integral is given by:

$$(30) \quad \int_\lambda^\infty e^{-\kappa^2 t^2 - |z-z_i|^2/(4t^2)} dt = \frac{\sqrt{\pi}}{4\kappa}\left[ e^{\kappa|z-z_i|}\mathrm{erfc}\left(\lambda\kappa + \frac{|z-z_i|}{2\lambda}\right) + e^{-\kappa|z-z_i|}\mathrm{erfc}\left(\lambda\kappa - \frac{|z-z_i|}{2\lambda}\right) \right]$$

And thus:

$$(31) \quad \begin{aligned} \varphi^L(\vec{\kappa} > 0, z) &= \sum_{i=1}^{d} \hat{\varphi}_i^L(\vec{\kappa}, z) = \frac{4\sqrt{\pi}}{T_x T_y}\sum_{i=1}^{4} q_i \int_\lambda^\infty e^{-\kappa^2 t^2 - |z-z_i|^2/(4t^2)} dt\, e^{-i\vec{\kappa}\cdot\vec{\rho}_i} = \\ &= \frac{\pi}{T_x T_y \kappa}\sum_{i=1}^{d}\left[ e^{\kappa|z-z_i|}\mathrm{erfc}\left(\lambda\kappa + \frac{|z-z_i|}{2\lambda}\right) + e^{-\kappa|z-z_i|}\mathrm{erfc}\left(\lambda\kappa - \frac{|z-z_i|}{2\lambda}\right)\right] q_i e^{-i\vec{\kappa}\cdot\vec{\rho}_i} \end{aligned}$$

### $\kappa = 0$

For $\kappa = 0$ the integral is divergent. Nevertheless, it can be written as a sum of a converging integral and a constant (not depending on i) divergent part as follows:

$$(32) \quad \begin{aligned} \int_\lambda^\infty e^{-\kappa^2 t^2 - |z-z_i|^2/(4t^2)} dt &\xrightarrow{\kappa=0} \int_\lambda^\infty e^{-|z-z_i|^2/(4t^2)} dt = \int_\lambda^\infty \left[ e^{-|z-z_i|^2/(4t^2)} - 1 \right] dt + \int_\lambda^\infty dt = \\ &= \lambda - \lambda e^{-\frac{|z-z_i|^2}{4\lambda^2}} - \frac{1}{2}\sqrt{\pi}|z - z_i|\mathrm{erf}\left(\frac{|z-z_i|}{2\lambda}\right) + \int_\lambda^\infty dt \end{aligned}$$

with this we obtain:

$$(33) \quad \begin{aligned} \varphi^L(\vec{\kappa}=0, z) &= \sum_{i=1}^{d} \hat{\varphi}_i^L(\vec{\kappa}=0, z) = \frac{4\sqrt{\pi}}{T_x T_y}\sum_{i=1}^{4} q_i \int_\lambda^\infty e^{-|z-z_i|^2/(4t^2)} dt = \\ &= \frac{4\sqrt{\pi}}{T_x T_y}\sum_{i=1}^{d} q_i\left[ \lambda - \lambda e^{-\frac{|z-z_i|^2}{4\lambda^2}} - \frac{1}{2}\sqrt{\pi}|z-z_i|\mathrm{erf}\left(\frac{|z-z_i|}{2\lambda}\right) + \int_\lambda^\infty dt \right] = \\ &= \frac{4\sqrt{\pi}}{T_x T_y}\sum_{i=1}^{d}\left[ -\lambda e^{-\frac{|z-z_i|^2}{4\lambda^2}} - \frac{1}{2}\sqrt{\pi}|z-z_i|\mathrm{erf}\left(\frac{|z-z_i|}{2\lambda}\right) \right] q_i + \sum_{i=1}^{4}\left[\left(\lambda + \int_\lambda^\infty dt\right) q_i\right] = \\ &= \frac{4\sqrt{\pi}}{T_x T_y}\sum_{i=1}^{d}\left[ -\lambda e^{-\frac{|z-z_i|^2}{4\lambda^2}} - \frac{1}{2}\sqrt{\pi}|z-z_i|\mathrm{erf}\left(\frac{|z-z_i|}{2\lambda}\right) \right] q_i + \left(\lambda + \int_\lambda^\infty dt\right)\sum_{i=1}^{4} q_i \end{aligned}$$

Since we treat charge neutral unit-cells such that $\sum_{i=1}^{d} q_i = 0$ (Eq. (1)) the divergent term falls and we are left with:

$$(34) \quad \varphi^L(\vec{\kappa}=0, z) = \frac{4\sqrt{\pi}}{T_x T_y} \sum_{i=1}^{d} \left[ -\lambda e^{-\frac{|z-z_i|^2}{4\lambda^2}} - \frac{1}{2}\sqrt{\pi}|z-z_i| \, erf\left(\frac{|z-z_i|}{2\lambda}\right) \right] q_i$$

We can now back-transform to real space:

$$(35) \quad \begin{aligned} \varphi^L(\vec{r}) &= \sum_{k_x} \sum_{k_y} \hat{\varphi}^L(\vec{\kappa}, z) e^{i\vec{\kappa}\cdot\vec{\rho}} = \frac{4\sqrt{\pi}}{T_x T_y} \sum_{i=1}^{d} \left[ -\lambda e^{-\frac{|z-z_i|^2}{4\lambda^2}} - \frac{1}{2}\sqrt{\pi}|z-z_i| \, erf\left(\frac{|z-z_i|}{2\lambda}\right) \right] q_i + \\ &+ \frac{\pi}{T_x T_y} \sum_{k_x} \sum_{k_y}^{*} \frac{1}{\sqrt{k_x^2 + k_y^2}} \times \\ &\sum_{i=1}^{d} \left[ e^{\sqrt{k_x^2+k_y^2}|z-z_i|} erfc\left(\lambda\sqrt{k_x^2+k_y^2} + \frac{|z-z_i|}{2\lambda}\right) + e^{-\sqrt{k_x^2+k_y^2}|z-z_i|} erfc\left(\lambda\sqrt{k_x^2+k_y^2} - \frac{|z-z_i|}{2\lambda}\right) \right] q_i e^{i[k_x(x-x_i) + k_y(y-y_i)]} \end{aligned}$$

Where $k_x = 0, \pm 2\pi/T_x, \pm 4\pi/T_x, \ldots$ and $k_y = 0, \pm 2\pi/T_y, \pm 4\pi/T_y, \ldots$ and the star sign indicates that we the term with $k_x = k_y = 0$ is excluded from the sum. This can be explicitly written as:

$$(36) \quad \begin{aligned} \varphi^L(\vec{r}) &= \frac{4\sqrt{\pi}}{T_x T_y} \sum_{i=1}^{d} \left[ -\lambda e^{-\frac{|z-z_i|^2}{4\lambda^2}} - \frac{1}{2}\sqrt{\pi}|z-z_i| \, erf\left(\frac{|z-z_i|}{2\lambda}\right) \right] q_i + \frac{1}{2} \sum_{n=-\infty}^{\infty} \sum_{m=-\infty}^{\infty}{}^{*} \frac{1}{\sqrt{n^2 T_y^2 + m^2 T_x^2}} \times \\ &\sum_{i=1}^{d} \left[ e^{2\pi\sqrt{(n/T_x)^2+(m/T_y)^2}|z-z_i|} erfc\left(2\pi\lambda\sqrt{(n/T_x)^2+(m/T_y)^2} + \frac{|z-z_i|}{2\lambda}\right) + \right. \\ &\left. e^{-2\pi\sqrt{(n/T_x)^2+(m/T_y)^2}|z-z_i|} erfc\left(2\pi\lambda\sqrt{(n/T_x)^2+(m/T_y)^2} - \frac{|z-z_i|}{2\lambda}\right) \right] q_i e^{2\pi i[(n/T_x)(x-x_i)+(m/T_y)(y-y_i)]} \end{aligned}$$

## The short range term

The short range term, $\varphi^S(\vec{r})$, is first back-transformed to real space (where it is absolutely convergent) and then summed.

$$\varphi^S(\vec{r}) = \sum_{k_x}\sum_{k_y} \hat{\varphi}^S(\vec{\kappa}, z)e^{i\vec{\kappa}\cdot\vec{\rho}} = \sum_{k_x}\sum_{k_y}\left[\frac{4\sqrt{\pi}}{T_xT_y}\sum_{i=1}^{d} q_i e^{-i\vec{\kappa}\cdot\vec{\rho}_i}\int_0^{\lambda} e^{-\kappa^2 t^2 - |z-z_i|^2/(4t^2)}dt\right]e^{i\vec{\kappa}\cdot\vec{\rho}} =$$

$$= \frac{4\sqrt{\pi}}{T_xT_y}\sum_{k_x}\sum_{k_y}\sum_{i=1}^{d} q_i e^{i\vec{\kappa}\cdot(\vec{\rho}-\vec{\rho}_i)}\int_0^{\lambda} e^{-\kappa^2 t^2 - |z-z_i|^2/(4t^2)}dt = \frac{4\sqrt{\pi}}{T_xT_y}\sum_{i=1}^{d} q_i \int_0^{\lambda}\left[\sum_{k_x}\sum_{k_y} e^{-\kappa^2 t^2}e^{i\vec{\kappa}\cdot(\vec{\rho}-\vec{\rho}_i)}\right]e^{-|z-z_i|^2/(4t^2)}dt =$$

$$(37) = \frac{4\sqrt{\pi}}{T_xT_y}\sum_{i=1}^{d} q_i \int_0^{\lambda}\left[\sum_{k_x}\sum_{k_y} e^{-(k_x^2+k_y^2)t^2}e^{i[k_x(x-x_i)+k_y(y-y_i)]}\right]e^{-|z-z_i|^2/(4t^2)}dt =$$

$$= 4\sqrt{\pi}\sum_{i=1}^{d} q_i \int_0^{\lambda}\left[\left(\frac{1}{T_x}\sum_{k_x} e^{-k_x^2 t^2}e^{ik_x(x-x_i)}\right)\left(\frac{1}{T_y}\sum_{k_y} e^{-k_y^2 t^2}e^{ik_y(y-y_i)}\right)\right]e^{-|z-z_i|^2/(4t^2)}dt$$

According to Poisson's summation formula (see proof below) we may write

$$(38)\quad \sum_{n=-\infty}^{\infty} f(x+nT) = \frac{1}{T}\sum_{m=-\infty}^{\infty} \hat{f}(m)e^{2\pi imx/T} \quad ;\quad \hat{f}(m) = \int_{-\infty}^{\infty} f(x)e^{-2\pi imx/T}dx$$

and applying to one of the sums above we obtain:

$$(39)\quad \frac{1}{T_x}\sum_{k_x} e^{-k_x^2 t^2}e^{ik_x(x-x_i)} = \frac{1}{T_x}\sum_{m=-\infty}^{\infty}\left[e^{-(2\pi n/T_x)^2 t^2}e^{-2\pi imx_i/T_x}\right]e^{2\pi imx/T_x}$$

where we can identify $\hat{f}(m) = e^{-(2\pi n/T_x)^2 t^2}e^{-2\pi imx_i/T_x}$. Owing to the following relation

$$\int_{-\infty}^{\infty}\left[\frac{1}{2\sqrt{\pi}t}e^{-(x-x_i)^2/(4t^2)}\right]e^{-2\pi imx/T_x}dx = e^{-(4\pi^2 m^2 t^2/T_x^2)}e^{-(2\pi imx_i/T_x)} = \hat{f}(m)\qquad \text{we find that}$$

$$f(x) = \frac{1}{2\sqrt{\pi}t}e^{-(x-x_i)^2/(4t^2)}.$$

Thus using Poisson's formula (Eq. (38)) we obtain:

$$(40)\quad \frac{1}{T_x}\sum_{k_x} e^{-k_x^2 t^2}e^{ik_x(x-x_i)} = \frac{1}{T_x}\sum_{m=-\infty}^{\infty}\left[e^{-(2\pi n/T_x)^2 t^2}e^{-2\pi imx_i/T_x}\right]e^{2\pi imx/T_x} = \sum_{n=-\infty}^{\infty}\frac{1}{2\sqrt{\pi}t}e^{-(x+nT_x-x_i)^2/(4t^2)}$$

Using this we can now write

$$\varphi^S(\vec{r}) = 4\sqrt{\pi}\sum_{i=1}^{d} q_i \int_0^\lambda \left[ \left( \frac{1}{T_x}\sum_{k_x} e^{-k_x^2 t^2} e^{ik_x(x-x_i)} \right) \left( \frac{1}{T_y}\sum_{k_y} e^{-k_y^2 t^2} e^{ik_y(y-y_i)} \right) \right] e^{-|z-z_i|^2/(4t^2)} dt =$$

$$= 4\sqrt{\pi}\sum_{i=1}^{d} q_i \int_0^\lambda \left[ \left( \sum_{n=-\infty}^{\infty} \frac{1}{2\sqrt{\pi}t} e^{-(x+nT_x-x_i)^2/(4t^2)} \right) \left( \sum_{m=-\infty}^{\infty} \frac{1}{2\sqrt{\pi}t} e^{-(y+mT_y-y_i)^2/(4t^2)} \right) \right] e^{-|z-z_i|^2/(4t^2)} dt =$$

(41)
$$= \frac{1}{\sqrt{\pi}} \sum_{i=1}^{d} q_i \sum_{n=-\infty}^{\infty} \sum_{m=-\infty}^{\infty} \int_0^\lambda \frac{e^{-[(x+nT_x-x_i)^2+(y+mT_y-y_i)^2+|z-z_i|^2]/(4t^2)}}{t^2} dt =$$

$$= \frac{1}{\sqrt{\pi}} \sum_{i=1}^{d} q_i \sum_{n=-\infty}^{\infty} \sum_{m=-\infty}^{\infty} \sqrt{\pi}\, \frac{\operatorname{erfc}\!\left(\sqrt{(x+nT_x-x_i)^2+(y+mT_y-y_i)^2+|z-z_i|^2}/(2\lambda)\right)}{\sqrt{(x+nT_x-x_i)^2+(y+mT_y-y_i)^2+|z-z_i|^2}} =$$

$$= \sum_{i=1}^{d} q_i \sum_{n=-\infty}^{\infty} \sum_{m=-\infty}^{\infty} \frac{\operatorname{erfc}\!\left(\sqrt{(x+nT_x-x_i)^2+(y+mT_y-y_i)^2+|z-z_i|^2}/(2\lambda)\right)}{\sqrt{(x+nT_x-x_i)^2+(y+mT_y-y_i)^2+|z-z_i|^2}}$$

So that

$$(42)\quad \boxed{\varphi^S(\vec{r}) = \sum_{i=1}^{d} q_i \sum_{n=-\infty}^{\infty} \sum_{m=-\infty}^{\infty} \frac{\operatorname{erfc}\!\left(\sqrt{(x+nT_x-x_i)^2+(y+mT_y-y_i)^2+|z-z_i|^2}/(2\lambda)\right)}{\sqrt{(x+nT_x-x_i)^2+(y+mT_y-y_i)^2+|z-z_i|^2}}}$$

# The Ewald method for hexagonal boron-nitride

Using the method described above, we can now write explicit expressions for calculating the electrostatic potential above a single layer of hexagonal boron nitride (*h*-BN). As explained above, we consider a rectangular unit-cell of the *h*-BN layer (see Fig. S1) with translational vectors $\vec{a}_x = \sqrt{3}a(1,0,0)$ and $\vec{a}_y = 3a(0,1,0)$, x-periodicity of $T_x = |\vec{a}_x| = \sqrt{3}a$ and y-periodicity of $T_y = |\vec{a}_y| = 3a$, $a$ being the B-N bond length.

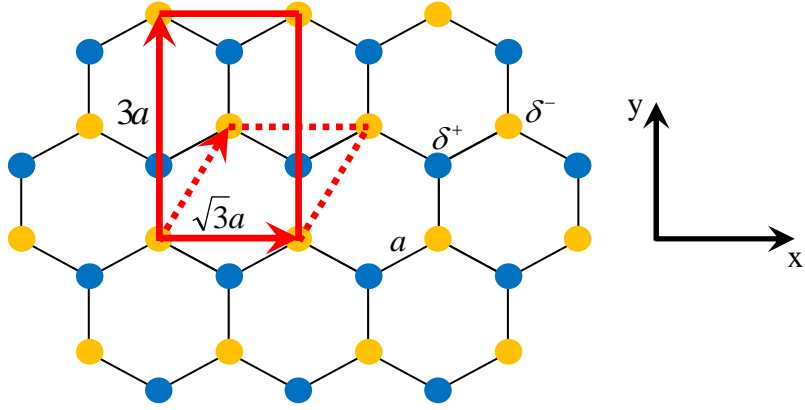

Fig. S6: Illustration of the unit cell used for the Ewald summation of the electrostatic potential above a two dimensional layer of pristine hexagonal boron nitride. Blue and orange circles represent boron and nitrogen atoms, respectively. Dashed red line shows the minimal (non-Cartesian) unit-cell consisting of two atoms while the full red line shows the rectangular Cartesian four atom unit-cell used in the calculations presented in the present study. $a$ and $\delta$ stand for the B-N bond length and partial charge on each atomic site, respectively (see main text for more details).

Without loss of generality we assume that the *h*-BN sheet is located in the XY-plane with Z=0 and calculate the electrostatic potential generated by the 2D sheet at a general point $\vec{r} = (x, y, z \neq 0)$. The rectangular unit cell includes $d = 4$ atoms located at the following positions:

(43) $\quad \vec{r}_1 = (0,0,0) \; ; \; \vec{r}_2 = (0,1,0)a \; ; \; \vec{r}_3 = \frac{1}{2}(\sqrt{3},3,0)a \; ; \; \vec{r}_4 = \frac{1}{2}(\sqrt{3},5,0)a$

and bearing the following charges:

(44) $\quad q_1 = \delta^- = -\delta \; ; \; q_2 = \delta^+ = +\delta \; ; \; q_3 = \delta^- = -\delta \; ; \; q_4 = \delta^+ = +\delta.$

such that it is charge neutral:

(45) $\quad \sum_{i=0}^{d} q_i = 0.$

**The long range term**

The long range term appearing in Eq. (36) is composed of two terms, one resulting from the case where $\kappa=0$ and the other with $\kappa>0$.

Since $z_i = 0$ (Eq. (43)) and the unit cell is neutral (Eq. (45)) the term with $\kappa=0$ vanishes

(46) $$\frac{4\sqrt{\pi}}{3\sqrt{3}a^2}\sum_{i=1}^{4}\left[-\lambda e^{-\frac{|z-z_i|^2}{4\lambda^2}} - \frac{1}{2}\sqrt{\pi}|z-z_i|\erf\left(\frac{|z-z_i|}{2\lambda}\right)\right]q_i = \frac{4\sqrt{\pi}}{3\sqrt{3}a^2}\left[-\lambda e^{-\frac{|z|^2}{4\lambda^2}} - \frac{1}{2}\sqrt{\pi}|z|\erf\left(\frac{|z|}{2\lambda}\right)\right]\sum_{i=1}^{4}q_i$$

Therefore, for the long range contribution to the potential we are left with the $\kappa>0$ term:

(47) $$\varphi^L(\vec{r}) = \frac{1}{2}\sum_{n=-\infty}^{\infty}\sum_{m=-\infty}^{\infty}{}^{*}\frac{1}{a\sqrt{9n^2+3m^2}}\left[e^{(2\pi/(3a))\sqrt{3n^2+m^2}|z|}\erfc\left((2\pi\lambda/(3a))\sqrt{3n^2+m^2}+\frac{|z|}{2\lambda}\right)+\right.$$
$$\left. e^{-(2\pi/(3a))\sqrt{3n^2+m^2}|z|}\erfc\left((2\pi\lambda/(3a))\sqrt{3n^2+m^2}-\frac{|z|}{2\lambda}\right)\right]\sum_{i=1}^{4}q_i e^{2\pi i[\sqrt{3}n(x-x_i)+m(y-y_i)]/(3a)}$$

This sum can be rewritten as a sum of three terms: (i) a term with $n = 0$ and $m \neq 0$, (ii) a term with $n \neq 0$ and $m = 0$, and (iii) a term with $n \neq 0$ and $m \neq 0$. Furthermore, we can limit the sums to the range $0,1,\ldots,\infty$ by collecting the terms with corresponding positive and negative indices: $\pm m$ and $\pm n$.

$\underline{m \neq n = 0}$

$$\varphi^L_{m \neq n=0}(\vec{r}) = \frac{1}{2}\sum_{m=-\infty}^{\infty}{}'\frac{1}{\sqrt{3}a|m|}\left[e^{(2\pi|m|/(3a))|z|}\erfc\left(2\pi|m|\lambda/(3a)+\frac{|z|}{2\lambda}\right)+e^{-(2\pi|m|/(3a))|z|}\erfc\left(2\pi|m|\lambda/(3a)-\frac{|z|}{2\lambda}\right)\right]\times$$
$$\sum_{i=1}^{4}q_i e^{2\pi im(y-y_i)/(3a)} =$$
$$= \frac{1}{2\sqrt{3}a}\sum_{m=1}^{\infty}\frac{1}{m}\left[e^{(2\pi m/(3a))|z|}\erfc\left((2\pi m\lambda/(3a))+\frac{|z|}{2\lambda}\right)+e^{-(2\pi m/(3a))|z|}\erfc\left((2\pi m\lambda/(3a))-\frac{|z|}{2\lambda}\right)\right]\times$$
$$\sum_{i=1}^{4}q_i\left[e^{2\pi im(y-y_i)/(3a)}+e^{-2\pi im(y-y_i)/(3a)}\right]=$$
$$= \frac{1}{\sqrt{3}a}\sum_{m=1}^{\infty}\frac{1}{m}\left[e^{(2\pi m/(3a))|z|}\erfc\left((2\pi m\lambda/(3a))+\frac{|z|}{2\lambda}\right)+e^{-(2\pi m/(3a))|z|}\erfc\left((2\pi m\lambda/(3a))-\frac{|z|}{2\lambda}\right)\right]\times$$
$$\sum_{i=1}^{4}q_i\cos[2\pi m(y-y_i)/(3a)]$$

So that

(48) $$\varphi^L_{m \neq n=0}(\vec{r}) = \frac{1}{\sqrt{3}a}\sum_{m=1}^{\infty}\frac{1}{m}\left[e^{(2\pi m/(3a))|z|}\erfc\left((2\pi m\lambda/(3a))+\frac{|z|}{2\lambda}\right)+e^{-(2\pi m/(3a))|z|}\erfc\left((2\pi m\lambda/(3a))-\frac{|z|}{2\lambda}\right)\right]\times$$
$$\sum_{i=1}^{4}q_i\cos[2\pi m(y-y_i)/(3a)]$$

$\underline{n \neq m = 0}$

$$\varphi^L_{n\neq m=0}(\vec{r}) = \frac{1}{2}\sum_{n=-\infty}^{\infty}{}'\frac{1}{3a|n|}\left[e^{(2\pi|n|/(\sqrt{3}a))|z|}\mathrm{erfc}\left((2\pi|n|\lambda/(\sqrt{3}a)) + \frac{|z|}{2\lambda}\right) + \right.$$

$$\left. + e^{-(2\pi|n|/(\sqrt{3}a))|z|}\mathrm{erfc}\left((2\pi|n|\lambda/(\sqrt{3}a)) - \frac{|z|}{2\lambda}\right)\right]\sum_{i=1}^{4}q_i e^{2\pi i n(x-x_i)/(\sqrt{3}a)} =$$

$$= \frac{1}{6a}\sum_{n=1}^{\infty}\frac{1}{n}\left[e^{(2\pi n/(\sqrt{3}a))|z|}\mathrm{erfc}\left((2\pi n\lambda/(\sqrt{3}a)) + \frac{|z|}{2\lambda}\right) + \right.$$

$$\left. + e^{-(2\pi n/(\sqrt{3}a))|z|}\mathrm{erfc}\left((2\pi n\lambda/(\sqrt{3}a)) - \frac{|z|}{2\lambda}\right)\right]\sum_{i=1}^{4}q_i\left[e^{2\pi i n(x-x_i)/(\sqrt{3}a)} + e^{-2\pi i n(x-x_i)/(\sqrt{3}a)}\right] =$$

$$= \frac{1}{3a}\sum_{n=1}^{\infty}\frac{1}{n}\left[e^{(2\pi n/(\sqrt{3}a))|z|}\mathrm{erfc}\left((2\pi n\lambda/(\sqrt{3}a)) + \frac{|z|}{2\lambda}\right) + \right.$$

$$\left. + e^{-(2\pi n/(\sqrt{3}a))|z|}\mathrm{erfc}\left((2\pi n\lambda/(\sqrt{3}a)) - \frac{|z|}{2\lambda}\right)\right]\sum_{i=1}^{4}q_i\cos(2\pi n(x-x_i)/(\sqrt{3}a))$$

So that

$$\varphi^L_{n\neq m=0}(\vec{r}) = \frac{1}{3a}\sum_{n=1}^{\infty}\frac{1}{n}\left[e^{(2\pi n/(\sqrt{3}a))|z|}\mathrm{erfc}\left((2\pi n\lambda/(\sqrt{3}a)) + \frac{|z|}{2\lambda}\right) + \right.$$

(49)
$$\left. + e^{-(2\pi n/(\sqrt{3}a))|z|}\mathrm{erfc}\left((2\pi n\lambda/(\sqrt{3}a)) - \frac{|z|}{2\lambda}\right)\right]\sum_{i=1}^{4}q_i\cos(2\pi n(x-x_i)/(\sqrt{3}a))$$

$\underline{n \neq 0\,;\, m \neq 0}$

$$\varphi^L(\vec{r}) = \frac{1}{2}\sum_{n=-\infty}^{\infty}\sum_{m=-\infty}^{\infty}{}^{*}\frac{1}{a\sqrt{9n^2+3m^2}}\left[e^{(2\pi/(3a))\sqrt{3n^2+m^2}|z|}\mathrm{erfc}\left((2\pi\lambda/(3a))\sqrt{3n^2+m^2} + \frac{|z|}{2\lambda}\right) + \right.$$

$$e^{-(2\pi/(3a))\sqrt{3n^2+m^2}|z|}\mathrm{erfc}\left((2\pi\lambda/(3a))\sqrt{3n^2+m^2} - \frac{|z|}{2\lambda}\right)\bigg]\sum_{i=1}^{4}q_i e^{2\pi i[\sqrt{3}n(x-x_i)+m(y-y_i)]/(3a)} =$$

$$= \frac{1}{2}\sum_{n=1}^{\infty}\sum_{m=1}^{\infty}\frac{1}{a\sqrt{9n^2+3m^2}}\left[e^{(2\pi/(3a))\sqrt{3n^2+m^2}|z|}\mathrm{erfc}\left((2\pi\lambda/(3a))\sqrt{3n^2+m^2} + \frac{|z|}{2\lambda}\right) + \right.$$

$$e^{-(2\pi/(3a))\sqrt{3n^2+m^2}|z|}\mathrm{erfc}\left((2\pi\lambda/(3a))\sqrt{3n^2+m^2} - \frac{|z|}{2\lambda}\right)\bigg] \times$$

$$\sum_{i=1}^{4}q_i\left[e^{2\pi i[\sqrt{3}n(x-x_i)+m(y-y_i)]/(3a)} + e^{-2\pi i[\sqrt{3}n(x-x_i)+m(y-y_i)]/(3a)} + e^{2\pi i[\sqrt{3}n(x-x_i)-m(y-y_i)]/(3a)} + e^{-2\pi i[\sqrt{3}n(x-x_i)-m(y-y_i)]/(3a)}\right] =$$

$$= \sum_{n=1}^{\infty}\sum_{m=1}^{\infty} \frac{1}{a\sqrt{9n^2+3m^2}} \left[ e^{(2\pi/(3a))\sqrt{3n^2+m^2}|z|} erfc\left( (2\pi\lambda/(3a))\sqrt{3n^2+m^2} + \frac{|z|}{2\lambda} \right) + \right.$$

$$\left. e^{-(2\pi/(3a))\sqrt{3n^2+m^2}|z|} erfc\left( (2\pi\lambda/(3a))\sqrt{3n^2+m^2} - \frac{|z|}{2\lambda} \right) \right] \times$$

$$\sum_{i=1}^{4} q_i \left[ \cos\left(2\pi[\sqrt{3}n(x-x_i)+m(y-y_i)]/(3a)\right) + \cos\left(2\pi[\sqrt{3}n(x-x_i)-m(y-y_i)]/(3a)\right) \right] =$$

$$= \frac{2}{a}\sum_{n=1}^{\infty}\sum_{m=1}^{\infty} \frac{1}{\sqrt{9n^2+3m^2}} \left[ e^{(2\pi/(3a))\sqrt{3n^2+m^2}|z|} erfc\left( (2\pi\lambda/(3a))\sqrt{3n^2+m^2} + \frac{|z|}{2\lambda} \right) + \right.$$

$$\left. e^{-(2\pi/(3a))\sqrt{3n^2+m^2}|z|} erfc\left( (2\pi\lambda/(3a))\sqrt{3n^2+m^2} - \frac{|z|}{2\lambda} \right) \right] \sum_{i=1}^{4} q_i \cos[2\pi n(x-x_i)/(\sqrt{3}a)] \cos[2\pi m(y-y_i)/(3a)]$$

so that

$$\varphi_{n\neq 0, m\neq 0}^{L}(\vec{r}) = \frac{2}{a}\sum_{n=1}^{\infty}\sum_{m=1}^{\infty} \frac{1}{\sqrt{9n^2+3m^2}} \left[ e^{(2\pi/(3a))\sqrt{3n^2+m^2}|z|} erfc\left( (2\pi\lambda/(3a))\sqrt{3n^2+m^2} + \frac{|z|}{2\lambda} \right) + \right.$$

(50) $$\left. e^{-(2\pi/(3a))\sqrt{3n^2+m^2}|z|} erfc\left( (2\pi\lambda/(3a))\sqrt{3n^2+m^2} - \frac{|z|}{2\lambda} \right) \right] \sum_{i=1}^{4} q_i \cos[2\pi n(x-x_i)/(\sqrt{3}a)] \cos[2\pi m(y-y_i)/(3a)]$$

Summing up all the long-range terms (Eqs. (48), (49), and (50)) we obtain:

(51)
$$\varphi^{L}(\vec{r}) = \frac{2}{a}\sum_{n=1}^{\infty}\sum_{m=1}^{\infty} \frac{1}{\sqrt{9n^2+3m^2}} \left[ e^{(2\pi/(3a))\sqrt{3n^2+m^2}|z|} erfc\left( (2\pi\lambda/(3a))\sqrt{3n^2+m^2} + \frac{|z|}{2\lambda} \right) + \right.$$

$$\left. e^{-(2\pi/(3a))\sqrt{3n^2+m^2}|z|} erfc\left( (2\pi\lambda/(3a))\sqrt{3n^2+m^2} - \frac{|z|}{2\lambda} \right) \right] \sum_{i=1}^{4} q_i \cos[2\pi n(x-x_i)/(\sqrt{3}a)] \cos[2\pi m(y-y_i)/(3a)] +$$

$$\frac{1}{\sqrt{3}a}\sum_{m=1}^{\infty} \frac{1}{m} \left[ e^{(2\pi m/(3a))|z|} erfc\left( (2\pi m\lambda/(3a)) + \frac{|z|}{2\lambda} \right) + e^{-(2\pi m/(3a))|z|} erfc\left( (2\pi m\lambda/(3a)) - \frac{|z|}{2\lambda} \right) \right] \times$$

$$\sum_{i=1}^{4} q_i \cos[2\pi n(y-y_i)/(3a)] +$$

$$\frac{1}{3a}\sum_{n=1}^{\infty} \frac{1}{n} \left[ e^{(2\pi n/(\sqrt{3}a))|z|} erfc\left( (2\pi n\lambda/(\sqrt{3}a)) + \frac{|z|}{2\lambda} \right) + e^{-(2\pi n/(\sqrt{3}a))|z|} erfc\left( (2\pi n\lambda/(\sqrt{3}a)) - \frac{|z|}{2\lambda} \right) \right] \times$$

$$\sum_{i=1}^{4} q_i \cos(2\pi n(x-x_i)/(\sqrt{3}a))$$

### The short range term

For the short range term we obtain (see Eq. 42):

$$(52) \quad \varphi^S(\vec{r}) = \sum_{i=1}^{4} q_i \sum_{n=-\infty}^{\infty} \sum_{m=-\infty}^{\infty} \frac{\text{erfc}\left(\sqrt{(x+n\sqrt{3}a-x_i)^2+(y+3ma-y_i)^2+z^2}\big/(2\lambda)\right)}{\sqrt{(x+n\sqrt{3}a-x_i)^2+(y+3ma-y_i)^2+z^2}}$$

## Proof of Poisson's summation formula (Eq. 38):

We define a function

$$(53) \quad h(x) = \sum_{n=-\infty}^{\infty} f(x+nT)$$

which is T-periodic $\left(h(x+mT) = \sum_{n=-\infty}^{\infty} f[x+(n+m)T]\underset{l\equiv n+m}{=}\sum_{l=-\infty}^{\infty} f(x+lT) = h(x)\right)$ and can therefore be expanded in a Fourier series such that:

$$(54) \quad h(x) = \sum_{n=-\infty}^{\infty} f(x+nT) = \sum_{m=-\infty}^{\infty} \hat{h}(m) e^{2\pi i m x/T}$$

With

$$(55) \quad \begin{aligned}\hat{h}(m) &= \frac{1}{T}\int_{-T/2}^{T/2} h(x)e^{-2\pi i m x/T}\,dx = \frac{1}{T}\int_{-T/2}^{T/2}\sum_{n=-\infty}^{\infty} f(x+nT)e^{-2\pi i m x/T}\,dx = \frac{1}{T}\sum_{n=-\infty}^{\infty}\int_{-T/2}^{T/2} f(x+nT)e^{-2\pi i m x/T}\,dx = \\ &\underset{\tilde{x}=x+nT}{=} \frac{1}{T}\sum_{n=-\infty}^{\infty}\int_{-T/2+nT}^{T/2+nT} f(\tilde{x})e^{-2\pi i m(\tilde{x}-nT)/T}\,d\tilde{x} = \frac{1}{T}\int_{-\infty}^{\infty} f(\tilde{x})e^{-2\pi i m\tilde{x}/T}e^{2\pi i m n}\,d\tilde{x} \underset{\tilde{x}=x}{=} \frac{1}{T}\int_{-\infty}^{\infty} f(x)e^{-2\pi i m x/T}\,dx\end{aligned}$$

Here, similar to Eq. (16) we have used the relation:

$$(56) \quad \sum_{n=-\infty}^{\infty}\int_{-T/2+nT}^{T/2+nT} d\tilde{x} = \cdots + \int_{-3T/2}^{-T/2} d\tilde{x} + \int_{-T/2}^{T/2} d\tilde{x} + \int_{T/2}^{3T/2} d\tilde{x} + \cdots = \int_{-\infty}^{\infty} d\tilde{x}$$

# The total monopolar Coulomb energy of infinite bilayer *h*-BN

Consider a *h*-BN bilayer of finite size consisting of *N* unit cells stacked in parallel exactly on top of each other in the AA' stacking mode with an interlayer distance *R*.

The intra-layer monopolar Coulomb interactions within each layer are independent of *R* and are therefore set to zero. We are interested in calculating the interlayer monopolar interaction energy given by (in atomic units):

$$E_N = \sum_{i=1}^{N} \sum_{d=1}^{2} \sum_{j=1}^{N} \sum_{e=1}^{2} \frac{q_{i,d}^{(1)} q_{j,e}^{(2)}}{r_{i,d;j,e}}$$

Where the sums over *i* and *j* run over all unit cells in layers 1 and 2, respectively, and the sums over *d* and *e* run over the two atoms within each unit cell of the corresponding layer. $q_{i,d}^{(1)}$ is the formal charge on atom *d* of unit cell *i* in layer 1, $q_{j,e}^{(2)}$ is the formal charge on atom *e* of unit cell *j* in layer 2, and $r_{i,d;j,e}$ is the distance between these two atoms. We may rewrite this sum as follows:

$$E_N = \sum_{i=1}^{N} \sum_{d=1}^{2} q_{i,d}^{(1)} \left( \sum_{j=1}^{N} \sum_{e=1}^{2} \frac{q_{j,e}^{(2)}}{r_{i,d;j,e}} \right) = \sum_{i=1}^{N} \sum_{d=1}^{2} q_{i,d}^{(1)} \phi_{i,d}^{(2)}$$

Where $\phi_{i,d}^{(2)} = \left( \sum_{j=1}^{N} \sum_{e=1}^{2} \frac{q_{j,e}^{(2)}}{r_{i,d;j,e}} \right)$ is the potential experienced by particle *d* in unit cell *i* of layer 1 due to all charges in layer two.

As the total energy may also be written as

$$E_N = \sum_{j=1}^{N} \sum_{e=1}^{2} \sum_{i=1}^{N} \sum_{d=1}^{2} \frac{q_{j,e}^{(2)} q_{i,d}^{(1)}}{r_{i,d;j,e}} = \sum_{j=1}^{N} \sum_{e=1}^{2} q_{j,e}^{(2)} \sum_{i=1}^{N} \sum_{d=1}^{2} \frac{q_{i,d}^{(1)}}{r_{i,d;j,e}} = \sum_{j=1}^{N} \sum_{e=1}^{2} q_{j,e}^{(2)} \phi_{j,e}^{(1)} \text{ with } \phi_{j,e}^{(1)} = \sum_{i=1}^{N} \sum_{d=1}^{2} \frac{q_{i,d}^{(1)}}{r_{j,e;i,d}}$$

($r_{i,d;j,e} = r_{j,e;i,d}$) one can write the total energy in a symmetrized manner as follows:

$$E_N = \frac{1}{2} \left[ \sum_{i=1}^{N} \sum_{d=1}^{2} q_{i,d}^{(1)} \phi_{i,d}^{(2)} + \sum_{j=1}^{N} \sum_{e=1}^{2} q_{j,e}^{(2)} \phi_{j,e}^{(1)} \right]$$

As the limit of an infinite bilayer will be taken it is desired to normalize the energy by the number of atoms. The term $\sum_{i=1}^{N} \sum_{d=1}^{2} q_{i,d}^{(1)} \phi_{i,d}^{(2)}$ corresponds to the energy contribution of the 2*N* atoms of the upper layer due to their interactions with the lower layer and the term $\sum_{j=1}^{N} \sum_{e=1}^{2} q_{j,e}^{(2)} \phi_{j,e}^{(1)}$ corresponds to the energy contribution of the 2*N* atoms of the lower layer due to their interaction with the upper layer. Thus, the total energy per atom is obtained by dividing the above terms by 2*N* as follows:

$$E_N = \frac{1}{2} \left[ \frac{1}{2N} \sum_{i=1}^{N} \sum_{d=1}^{2} q_{i,d}^{(1)} \phi_{i,d}^{(2)} + \frac{1}{2N} \sum_{j=1}^{N} \sum_{e=1}^{2} q_{j,e}^{(2)} \phi_{j,e}^{(1)} \right] = \frac{1}{4N} \left[ \sum_{i=1}^{N} \sum_{d=1}^{2} q_{i,d}^{(1)} \phi_{i,d}^{(2)} + \sum_{j=1}^{N} \sum_{e=1}^{2} q_{j,e}^{(2)} \phi_{j,e}^{(1)} \right]$$

In the above expression $\sum_{d=1}^{2} q_{i,d}^{(1)} \phi_{i,d}^{(2)}$ is the interaction energy of unit cell $i$ in layer 1 with all the charges in layer 2. When the limit $N \to \infty$ is taken all unit cells become equivalent and therefore one may replace the general cell index $i$ by any cell index, say 1, and write $\sum_{d=1}^{2} q_{i,d}^{(1)} \phi_{i,d}^{(2)} \xrightarrow{N \to \infty} \sum_{d=1}^{2} q_{1,d}^{(1)} \phi_{1,d}^{(2)}$. Similarly, one finds that $\sum_{e=1}^{2} q_{j,e}^{(2)} \phi_{j,e}^{(1)} \xrightarrow{N \to \infty} \sum_{e=1}^{2} q_{1,e}^{(2)} \phi_{1,e}^{(1)}$. With this we may write the total energy per atom in the limit of an infinite bilayer as:

$$E_{N \to \infty} = \frac{1}{4N}\left[\sum_{i=1}^{N}\sum_{d=1}^{2} q_{i,d}^{(1)}\phi_{i,d}^{(2)} + \sum_{j=1}^{N}\sum_{e=1}^{2} q_{j,e}^{(2)}\phi_{j,e}^{(1)}\right] = \frac{1}{4N}\left[\sum_{i=1}^{N}\sum_{d=1}^{2} q_{1,d}^{(1)}\phi_{1,d}^{(2)} + \sum_{j=1}^{N}\sum_{e=1}^{2} q_{1,e}^{(2)}\phi_{1,e}^{(1)}\right] =$$

$$= \frac{1}{4N}\left[\sum_{d=1}^{2} q_{1,d}^{(1)}\phi_{1,d}^{(2)}\sum_{i=1}^{N} 1 + \sum_{e=1}^{2} q_{1,e}^{(2)}\phi_{1,e}^{(1)}\sum_{j=1}^{N} 1\right] = \frac{1}{4N}\left[N\sum_{d=1}^{2} q_{1,d}^{(1)}\phi_{1,d}^{(2)} + N\sum_{e=1}^{2} q_{1,e}^{(2)}\phi_{1,e}^{(1)}\right] =$$

$$= \frac{1}{4}\left[\sum_{d=1}^{2} q_{1,d}^{(1)}\phi_{1,d}^{(2)} + \sum_{e=1}^{2} q_{1,e}^{(2)}\phi_{1,e}^{(1)}\right] = \frac{1}{4}\left[\sum_{d=1}^{2} q_{1,d}^{(1)}\sum_{j=1}^{N}\sum_{e=1}^{2} \frac{q_{j,e}^{(2)}}{r_{1,d;j,e}} + \sum_{e=1}^{2} q_{1,e}^{(2)}\sum_{i=1}^{N}\sum_{d=1}^{2} \frac{q_{i,d}^{(1)}}{r_{i,d;1,e}}\right] =$$

$$= \frac{1}{4}\left[\sum_{d=1}^{2}\sum_{j=1}^{N}\sum_{e=1}^{2} \frac{q_{1,d}^{(1)}q_{j,e}^{(2)}}{r_{1,d;j,e}} + \sum_{e=1}^{2}\sum_{i=1}^{N}\sum_{d=1}^{2} \frac{q_{1,e}^{(2)}q_{i,d}^{(1)}}{r_{i,d;1,e}}\right]$$

As the two layers in $h$-BN are equivalent the term $\sum_{d=1}^{2} q_{1,d}^{(1)}\phi_{1,d}^{(2)}$ representing monopolar electrostatic energy of the interaction of one unit cell in layer 1 with all atoms in layer 2 is equal to the monopolar electrostatic energy of the interaction of one unit cell in layer 2 with all atoms in layer 1 given by $\sum_{e=1}^{2} q_{1,e}^{(2)}\phi_{1,e}^{(1)}$. Thus, the total energy per atom of the infinite bilayer system is given by:

$$E_{\infty}^{coul} = \frac{1}{2}\sum_{d=1}^{2}\sum_{j=1}^{\infty}\sum_{e=1}^{2} \frac{q_{1,d}^{(1)}q_{j,e}^{(2)}}{r_{1,d;j,e}}$$

Furthermore, due to the lattice symmetry of the AA' stacked $h$-BN bilayer the potential experience by the two atoms within each unit cell is equal in magnitude and opposite in sign. Since the formal charges of the two atoms in the unit cell are opposite the monopolar electrostatic energy they contribute is equal resulting in the following final expression for the total energy per atom of the infinite bilayer $h$-BN system:

$$E_{\infty}^{coul} = \frac{1}{2}\sum_{d=1}^{2}\sum_{j=1}^{\infty}\sum_{e=1}^{2} \frac{q_{1,d}^{(1)}q_{j,e}^{(2)}}{r_{1,d;j,e}} = \frac{1}{2} \cdot 2\sum_{j=1}^{\infty}\sum_{e=1}^{2} \frac{q_{1,1}^{(1)}q_{j,e}^{(2)}}{r_{1,1;j,e}} = \sum_{j=1}^{\infty}\sum_{e=1}^{2} \frac{q_{1,1}^{(1)}q_{j,e}^{(2)}}{r_{1,1;j,e}} = q_{1,1}^{(1)}\sum_{j=1}^{\infty}\sum_{e=1}^{2} \frac{q_{j,e}^{(2)}}{r_{1,1;j,e}} = q_{1,1}^{(1)}\phi_{1,1}^{(2)}$$

or:

$$\boxed{E_{\infty}^{coul} = q_{1,1}^{(1)}\sum_{j=1}^{\infty}\sum_{e=1}^{2} \frac{q_{j,e}^{(2)}}{r_{1,1;j,e}}}$$

Namely, the total monopolar electrostatic energy per atom of the infinite bilayer $h$-BN system equals the energy of a charge placed above a lattice site of a single infinite $h$-BN layer.

# The total vdW energy of infinite bilayer graphene and *h*-BN

Consider a finite sized *h*-BN or graphene bilayer consisting of $N$ unit cells stacked in parallel exactly on top of each other in the appropriate stacking mode with an interlayer distance $R$.

The intra-layer vdW interactions within each layer are independent of $R$ and are therefore set to zero. We are interested in calculating the interlayer vdW energy given by:

$$E_N = \sum_{i=1}^{N} \sum_{d=1}^{2} \sum_{j=1}^{N} \sum_{e=1}^{2} \frac{c_6^{i,d;j,e}}{r_{i,d;j,e}^6}$$

Where the sums over $i$ and $j$ run over all unit cells in layers 1 and 2, respectively, and the sums over $d$ and $e$ run over the two atoms within each unit cell of the corresponding layer. $c_6^{i,d;j,e}$ is the appropriate $c_6$ coefficient between atom $d$ of unit cell $i$ in layer 1 and atom $e$ of unit cell $j$ in layer 2 and $r_{i,d;j,e}$ is the distance between these two atoms. One may rewrite the energy expression as:

$$E_N = \frac{1}{2}\left[\sum_{i=1}^{N}\sum_{d=1}^{2}\left(\sum_{j=1}^{N}\sum_{e=1}^{2}\frac{c_6^{i,d;j,e}}{r_{i,d;j,e}^6}\right) + \sum_{j=1}^{N}\sum_{e=1}^{2}\left(\sum_{i=1}^{N}\sum_{d=1}^{2}\frac{c_6^{i,d;j,e}}{r_{i,d;j,e}^6}\right)\right]$$

Where the term $\sum_{j=1}^{N}\sum_{e=1}^{2}\frac{c_6^{i,d;j,e}}{r_{i,d;j,e}^6}$ is the vdW energy contribution of atom $d$ in unit cell $i$ of layer 1 due to its interaction with all the atoms in layer 2 and the term $\sum_{i=1}^{N}\sum_{d=1}^{2}\frac{c_6^{i,d;j,e}}{r_{i,d;j,e}^6} = \sum_{i=1}^{N}\sum_{d=1}^{2}\frac{c_6^{j,e;i,d}}{r_{j,e;i,d}^6}$ is the vdW energy contribution of atom $e$ in unit cell $j$ of layer 2 due to its interaction with all the atoms in layer 1.

As the limit of an infinite bilayer will be taken it is desired to normalize the energy by the number of atoms. Thus, the two terms in the energy expression representing the sum of all individual contributions from each layer should be divided by the number of atoms in the layer, $2N$. Thus, the total energy per atom is given by:

$$E_N = \frac{1}{2}\left[\frac{1}{2N}\sum_{i=1}^{N}\sum_{d=1}^{2}\left(\sum_{j=1}^{N}\sum_{e=1}^{2}\frac{c_6^{i,d;j,e}}{r_{i,d;j,e}^6}\right) + \frac{1}{2N}\sum_{j=1}^{N}\sum_{e=1}^{2}\left(\sum_{i=1}^{N}\sum_{d=1}^{2}\frac{c_6^{i,d;j,e}}{r_{i,d;j,e}^6}\right)\right] = \frac{1}{2N}\sum_{i=1}^{N}\sum_{d=1}^{2}\left(\sum_{j=1}^{N}\sum_{e=1}^{2}\frac{c_6^{i,d;j,e}}{r_{i,d;j,e}^6}\right)$$

In the above expression the term $\sum_{d=1}^{2}\left(\sum_{j=1}^{N}\sum_{e=1}^{2}\frac{c_6^{i,d;j,e}}{r_{i,d;j,e}^6}\right)$ is the vdW interaction energy of unit cell $i$ in layer 1 with all the atoms in layer 2. When the limit $N \to \infty$ is taken all unit cells become equivalent and therefore one may replace the general cell index $i$ by any cell index, say 1, and write $\sum_{d=1}^{2}\left(\sum_{j=1}^{N}\sum_{e=1}^{2}\frac{c_6^{i,d;j,e}}{r_{i,d;j,e}^6}\right) \xrightarrow{N \to \infty} \sum_{d=1}^{2}\left(\sum_{j=1}^{N}\sum_{e=1}^{2}\frac{c_6^{1,d;j,e}}{r_{1,d;j,e}^6}\right)$. With this we may write the total energy per atom in the limit of an infinite bilayer as:

$$E_{N\to\infty} = \frac{1}{2N}\sum_{i=1}^{N}\sum_{d=1}^{2}\left(\sum_{j=1}^{N}\sum_{e=1}^{2}\frac{c_6^{i,d;j,e}}{r_{i,d;j,e}^6}\right) = \frac{1}{2N}\sum_{i=1}^{N}\sum_{d=1}^{2}\left(\sum_{j=1}^{N}\sum_{e=1}^{2}\frac{c_6^{1,d;j,e}}{r_{1,d;j,e}^6}\right) = \frac{1}{2N}\sum_{d=1}^{2}\left(\sum_{j=1}^{N}\sum_{e=1}^{2}\frac{c_6^{1,d;j,e}}{r_{1,d;j,e}^6}\right)\sum_{i=1}^{N}1 =$$

$$= \frac{1}{2N}\cdot N\sum_{d=1}^{2}\left(\sum_{j=1}^{N}\sum_{e=1}^{2}\frac{c_6^{1,d;j,e}}{r_{1,d;j,e}^6}\right) = \frac{1}{2}\sum_{d=1}^{2}\left(\sum_{j=1}^{N}\sum_{e=1}^{2}\frac{c_6^{1,d;j,e}}{r_{1,d;j,e}^6}\right)$$

or:

$$\boxed{E_\infty^{vdW} = \frac{1}{2}\sum_{d=1}^{2}\left(\sum_{j=1}^{N}\sum_{e=1}^{2}\frac{c_6^{1,d;j,e}}{r_{1,d;j,e}^6}\right)}$$

Namely, the total vdW energy per atom of the infinite bilayer *h*-BN or graphene systems equals half the vdW energy of a single unit cell placed above a single infinite layer of the corresponding material.